%% file: main.tex
\documentclass[lettersize,journal]{IEEEtran}
\usepackage{amsmath}
\usepackage{amsfonts}
\usepackage{subcaption}
\usepackage{algorithmic}
\usepackage{algorithm}
\usepackage{array}
\usepackage[caption=false,font=normalsize,labelfont=sf,textfont=sf]{subfig}
\usepackage{textcomp}
\usepackage{stfloats}
\usepackage{url}
\usepackage{verbatim}
\usepackage{graphicx}
\usepackage{cite}
\usepackage{flushend}
\usepackage{caption}
\usepackage{tabularx}
\usepackage{subcaption}
\usepackage{amsthm}
\usepackage{multirow}
\usepackage{color}
\usepackage{xcolor}
\usepackage{booktabs}
\usepackage{float}
\usepackage{xspace}
\usepackage{threeparttable}
\usepackage{url}
\usepackage{enumitem}
\usepackage{amssymb}
\usepackage{flushend}

\newcommand{\ie}{\textit{i}.\textit{e}.}
\newcommand{\eg}{\textit{e}.\textit{g}.} 
\newcommand{\wrt}{\textit{w}.\textit{r}.\textit{t} } 
\newtheorem{Def}{Definition}

\newtheorem*{Pro*}{Problem}
\newcommand{\model}{DCMGNN\xspace}

\begin{document}

\title{Dual-Channel Multiplex Graph Neural Networks for Recommendation}

\author{Xiang~Li,
        Chaofan~Fu,
        Zhongying Zhao,~\IEEEmembership{Member,~IEEE,}
        Guanjie Zheng,~\IEEEmembership{Member,~IEEE,}
        Chao Huang,~\IEEEmembership{Member,~IEEE,}
        Yanwei~Yu*,~\IEEEmembership{Member,~IEEE,}
        and~Junyu~Dong,~\IEEEmembership{Member,~IEEE}
\thanks{X. Li, C. Fu, Y. Yu, and J. Dong are with the Faculty of Information Science and Engineering, Ocean University of China.}
\thanks{Z. Zhao is with the College of Computer Science and Engineering, Shandong University of Science and Technology.}
\thanks{G. Zheng is with the Department of Computer Science and Engineering, Shanghai Jiao Tong University.}
\thanks{C. Huang is with the Department of Computer Science, The University of Hong Kong.}
\thanks{Y. Yu is the corresponding author.}
}

\markboth{IEEE TRANSACTIONS ON KNOWLEDGE and Data ENGINEERING, VOL.~37, NO.~6, June~2025}%
{Shell \MakeLowercase{\textit{et al.}}: A Sample Article Using IEEEtran.cls for IEEE Journals}

\maketitle

\begin{abstract}
Effective recommender systems play a crucial role in accurately capturing user and item attributes that mirror individual preferences. Some existing recommendation techniques have started to shift their focus towards modeling various types of interactive relations between users and items in real-world recommendation scenarios, such as clicks, marking favorites, and purchases on online shopping platforms. Nevertheless, these approaches still grapple with two significant challenges: (1) Insufficient modeling and exploitation of the impact of various behavior patterns formed by multiplex relations between users and items on representation learning, and (2) ignoring the effect of different relations within behavior patterns on the target relation in recommender system scenarios. In this work, we introduce a novel recommendation framework, \textbf{\underline{D}}ual-\textbf{\underline{C}}hannel \textbf{\underline{M}}ultiplex \textbf{\underline{G}}raph \textbf{\underline{N}}eural \textbf{\underline{N}}etwork (DCMGNN), which addresses the aforementioned challenges. It incorporates an explicit behavior pattern representation learner to capture the behavior patterns composed of multiplex user-item interactive relations, and includes a relation chain representation learner and a relation chain-aware encoder to discover the impact of various auxiliary relations on the target relation, the dependencies between different relations, and mine the appropriate order of relations in a behavior pattern. Extensive experiments on three real-world datasets demonstrate that our \model surpasses various state-of-the-art recommendation methods. It outperforms the best baselines by 10.06\% and 12.15\% on average across all datasets in terms of Recall@10 and NDCG@10, respectively.
\end{abstract}


\input{Main/1_Introduction}
\input{Main/2_Related_Work}
\input{Main/3_Preliminary}
\input{Main/4_Method}
\input{Main/5_Experiments}
\input{Main/6_Conclusion}

\section*{acknowledge}
This work is partially supported by the National Natural Science Foundation of China under grant Nos. 62176243, 62472263, and 41927805.

\vspace{-2mm}
\bibliographystyle{IEEEtran}
\bibliography{IEEE}

\vspace{-33pt}
\begin{IEEEbiography}[{\includegraphics[width=1in,height=1.25in,clip,keepaspectratio]{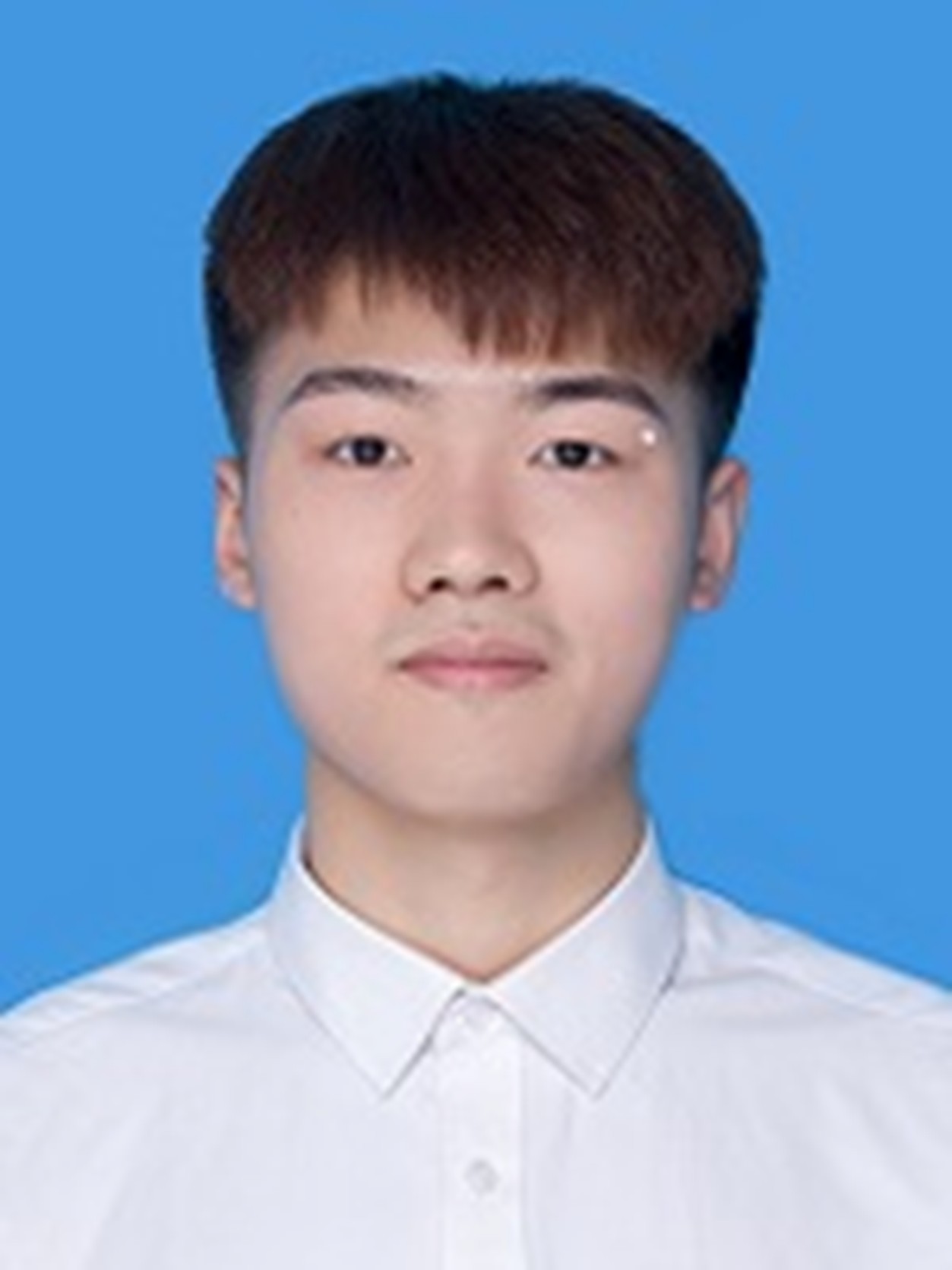}}]{Xiang Li}
\textnormal{is currently pursuing a Ph.D. degree with the Faculty of Information Science and Engineering, Ocean University of China, Qingdao, China. He is engaged in the research of graph neural networks, recommender systems, and data mining.}
\end{IEEEbiography}

\vspace{-33pt}
\begin{IEEEbiography}[{\includegraphics[width=1in,height=1.25in,clip,keepaspectratio]{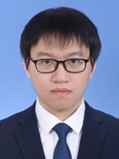}}]{Chaofan Fu}
 \textnormal{received the M.S. degree from the Faculty of Information Science and Engineering, Ocean University of China, Qingdao, China in 2024. His major research interests include the graph neural network and data mining.}
\end{IEEEbiography}

\vspace{-33pt}
\begin{IEEEbiography}[{\includegraphics[width=1in,height=1.25in,clip,keepaspectratio]{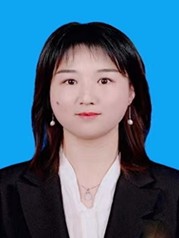}}]{Zhongying Zhao}
\textnormal{(Member, IEEE) received the Ph.D. degree from the Institute of Computing Technology, Chinese Academy of Sciences in 2012. From 2012 to 2014, she was an assistant professor in Shenzhen Institutes of Advanced Technology, Chinese Academy of Sciences. She is currently a professor with the College of Computer Science and Engineering, Shandong University of Science and Technology. Her research interests include social network analysis, graph neural networks, and data mining. }
\end{IEEEbiography}

\vspace{-33pt}
\begin{IEEEbiography}[{\includegraphics[width=1in,height=1.25in,clip,keepaspectratio]{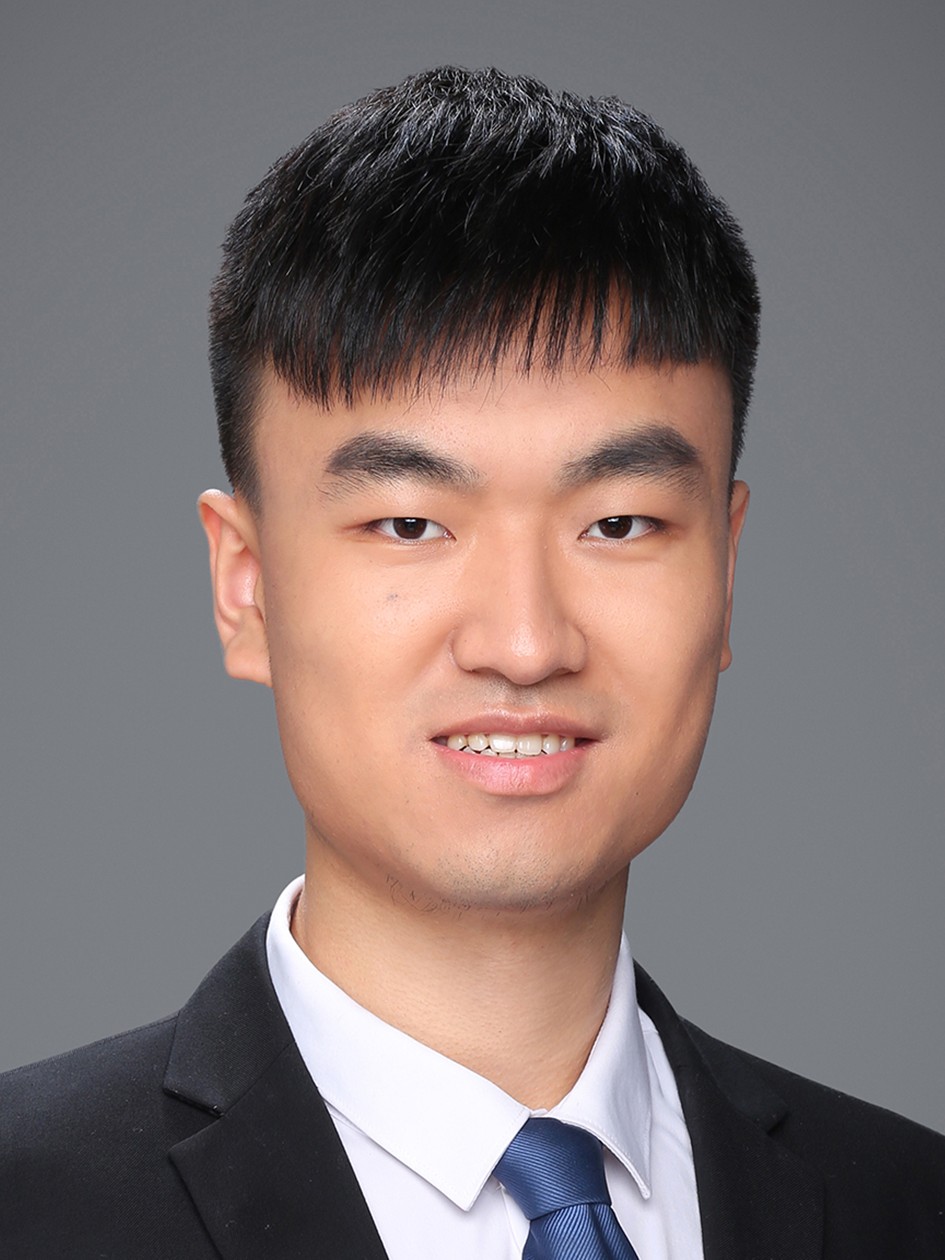}}]{Guanjie Zheng}
\textnormal{(Member, IEEE). He received the Ph.D. degree from the College of Information Sciences and Technology, The Pennsylvania State University, in 2020. Before that, he received the Bachelor degree in Electrical Engineering from Shanghai Jiao Tong University in 2015. He is currently an assistant professor in John Hopcroft Center for Computer Science at Shanghai Jiao Tong University. His research interest is data-driven intelligent decision-making, and its application on spatial temporal data.}
\end{IEEEbiography}

\vspace{-33pt}
\begin{IEEEbiography}[{\includegraphics[width=1in,height=1.25in,clip,keepaspectratio]{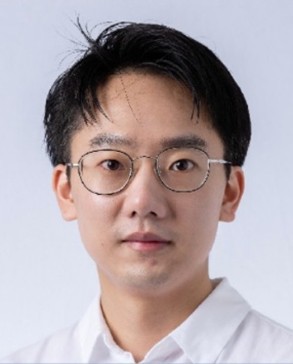}}]{Chao Huang}
 \textnormal{(Member, IEEE) received the Ph.D. degree from the University of Notre Dame. He is a tenure-track assistant professor with the Department of Computer Science and Musketeers Foundation Institute of Data Science, University of Hong Kong, Hong Kong. His research interests include applied machine learning, graph neural networks, recommendation, and spatial-temporal data mining.
 }
\end{IEEEbiography}

\vspace{-33pt}
\begin{IEEEbiography}[{\includegraphics[width=1in,height=1.25in,clip,keepaspectratio]{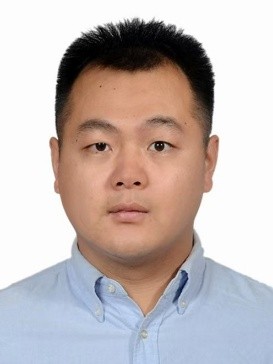}}]{Yanwei Yu}
\textnormal{(Member, IEEE) received the Ph.D. degree in computer science from the University of Science and Technology Beijing, Beijing, China, in 2014. 
From 2012 to 2013, he was a visiting scholar with the Worcester Polytechnic Institute. From 2016 to 2018, he was a postdoctoral researcher with the College of Information Sciences and Technology, Pennsylvania State University. He is currently a Professor at the College of Computer Science and Technology, Ocean University of China. His research interests include data mining and machine learning.}
\end{IEEEbiography}

\vspace{-33pt}
\begin{IEEEbiography}[{\includegraphics[width=1in,height=1.25in,clip,keepaspectratio]{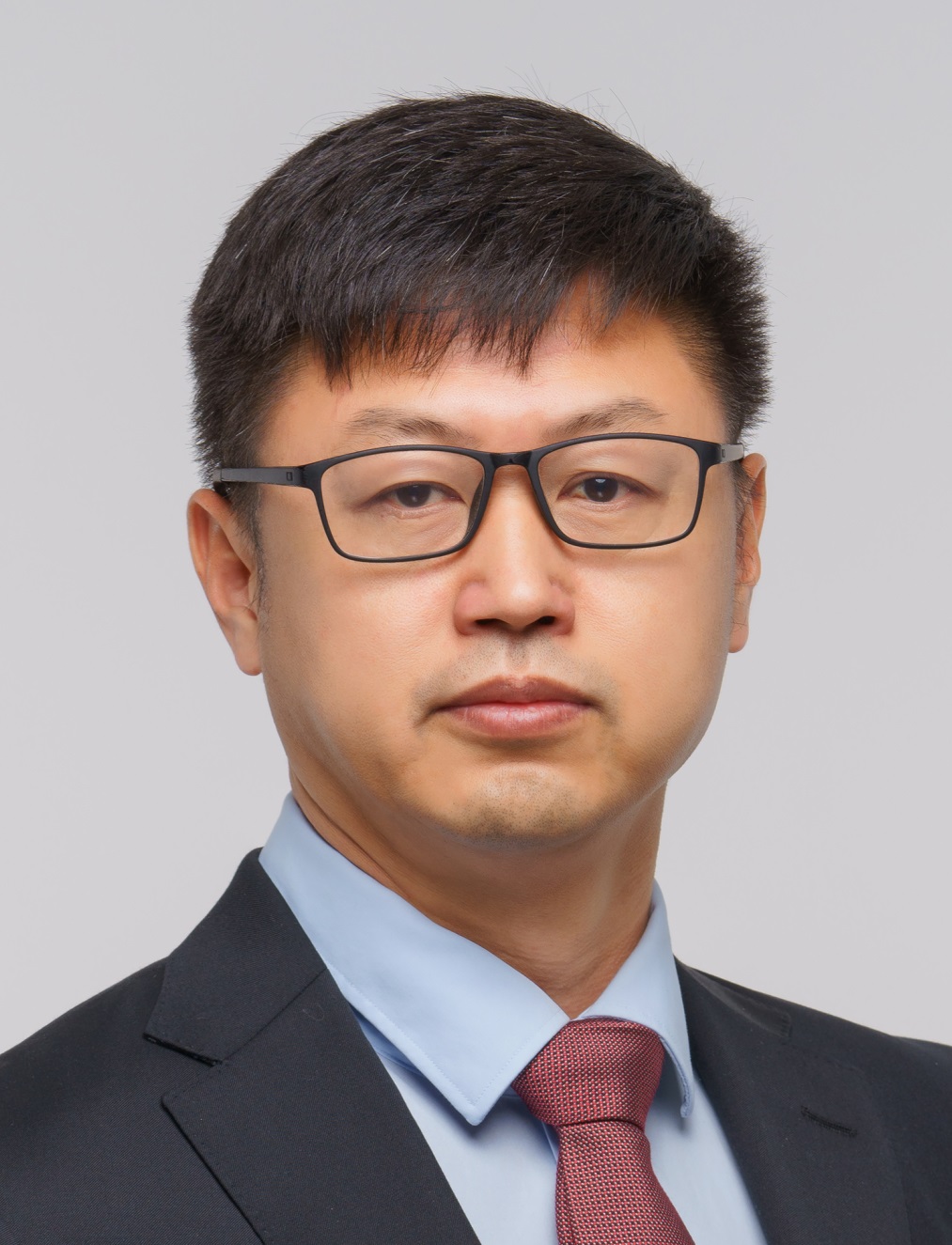}}]{Junyu Dong}
\textnormal{(Member, IEEE) received the B.Sc. and M.Sc. degrees from the Department of Applied Mathematics, Ocean University of China, Qingdao, China, in 1993 and 1999, respectively, and the Ph.D. degree in image processing from the Department of Computer Science, Heriot-Watt University, Edinburgh, U.K., in 2003. 
He is currently a Professor and the Dean with the Faculty of Information Science and Engineering, Ocean University of China. His research interests include computer vision and machine learning.}
\end{IEEEbiography}

\end{document}

%% file: Main/1_Introduction.tex
\vspace{-2mm}
\section{Introduction}
\IEEEPARstart{R}{ecommender} systems have emerged as essential elements across numerous online platforms, particularly in the realm of e-commerce~\cite{guo2019buying,deldjoo2020recommender,shen2022mbn,shui2024hierarchical,zhang2025learning}, social media~\cite{anandhan2018social}, and video platforms~\cite{cheuque2019recommender}, to deliver personalized and precise information recommendations to users~\cite{ren2023sslrec}. The interactions between users and items have grown increasingly diverse and complex~\cite{rendle2012bpr}, and traditional collaborative filtering (CF)-based recommendation~\cite{koren2009matrix,cheng2018aspect} tend to simplify the relations between users and items and are unable to fully reflect the diversity of user behaviors and deep interaction patterns. The advancement of graph neural networks (GNNs) has led to the mainstream adoption of constructing diverse GNNs to capture higher-order node neighborhood information~\cite{wu2022survey}.

Some existing GNN-based recommendation methods, such as HCCF~\cite{xia2022hypergraph}, DCCF~\cite{ren2023disentangled}, LightGCN \cite{he2020lightgcn}, LightGCL~\cite{cai2023lightgcl}, have not been limited to historical collaborative filtering information~\cite{he2017neural,mao2021ultragcn,cheng2022feature}, but enhance the representations of user-item interaction information by introducing techniques such as hypergraph structures~\cite{xia2022hypergraph}, singular value decomposition~\cite{cai2023lightgcl}, generative self-supervised learning~\cite{xia2023automated}, knowledge-aware learning~\cite{xia2021knowledge,chang2023meta,meng2023parallel}. Although existing methods have shown promising results, they typically concentrate on a single user-item interactive relation (generally called the target relation, \eg, purchase or buy relation in e-commerce networks), and do not fully capture the complicated interactive relations between users and items. To solve the problem, some researchers have proposed multi-behavior recommender systems~\cite{schlichtkrull2018modeling,gao2019learning,jin2020multi,gan2023multi,yan2024behavior}, which attempt to meticulously consider multiplex relations between users and items by introducing meta-paths~\cite{yang2021hyper}, contrastive learning~\cite{wei2022contrastive,gu2022self,wei2023multi,li2023intra,wu2022multi,lan2024contrastive}, graph disentanglement~\cite{ren2023disentangled}, and other means~\cite{yan2023cascading,cheng2023multi,meng2023hierarchical,xuan2023temporal,zhang2023denoising}. 

Multi-behavior recommendation task aims at predicting whether the user and the item will have a target relation based on the auxiliary relations, which requires the model to have the capability to capture the correlations between auxiliary relations and the target relation. However, existing approaches often struggle to fully capture the intricate influence of user-item interactions, shaped by multiplex relations, on their respective representations. In response, a specific group of researchers has aimed to address this limitation by capturing the behaviors through the development of multiplex graph neural networks~\cite{xia2020multiplex,wu2022graph}, such as HybridGNN~\cite{gu2022hybridgnn}, MHGCN~\cite{yu2022multiplex} and BPHGNN~\cite{fu2023multiplex}. While these models explicitly account for behavior patterns within multiplex relations, in recommender system scenarios they frequently disregard how various auxiliary relations in behavior patterns impact the target relation.

Although most current GNN-based recommender systems consider the influence of multiple interactive relations between users and items, they face two key limitations. \textit{First, the impact of various behavior patterns formed by multiplex interactive relations in heterogeneous graphs on recommendation is not adequately explored and exploited in existing recommender systems.} Real-world user-item interactions are complex and require nuanced representation learning to capture diverse user behaviors and evolving preferences~\cite{gan2023multi, xin2023improving, gao2023survey}. Unfortunately, existing general GNN architectures face significant limitations in effectively capturing meaningful multiplex node representations derived from complex multi-behavior relationships for recommendation tasks. We observe that multiple relations naturally form a multiplex heterogeneous network structure, making it critical to explicitly model multi-behavior interactions (\ie, behavior patterns) and learn representations of users. For instance, in the toy example of Fig.~\ref{fig:framework}, the multi-behavior interaction (`View', `Cart', `Buy') between user $u_1$ and item $v_1$ differs from (`View', `Buy') between $u_4$ and $v_5$ due to the absence of the `Cart' relation. For $u_1$ and $v_1$, `Cart' serves as a transitional step, linking `View' and `Buy' and reflecting a progressive strengthening of purchase intention. Conversely, for $u_4$ and $v_5$, the absence of `Cart' highlights the direct influence of `View' on `Buy'. These examples demonstrate that multi-behavior interactions cannot be treated as simple linear combinations of single-layer interactions. Instead, explicit behavior pattern modeling of the interplay of these relationships is crucial, as traditional methods often overlook such distinctions, leading to misinterpretation of user intent.

\textit{Second, most existing approaches ignore the impact of different interactive relations within behavior patterns on the target relation.} User behaviors are inherently multiplex, and diverse interactive relations within user behaviors influence the recommender system's ability to predict the target relation differently. 

By way of illustration, in the multi-behavior interaction between user $u_1$ and item $v_1$, besides the target relation `Buy', there exist two auxiliary relations: `View' and `Cart'. Crucially, the influences of auxiliary relations on `Buy' are not simply additive. While the most frequent relation `View' reflects initial purchase interest, the less frequent `Cart' indicates a more deliberate purchase intent. Another example is the multi-behavior interaction between $u_4$ and $v_5$, despite just lacking a `Cart' relation, `View' directly influences `Buy' without a transitional relation, altering the sequential dependencies and correlations among relations. Most existing methods overlook the influences of different auxiliary relations on the target relation within multi-behavior interactions, and the sequential correlations and dependencies among relations, leading to inaccurate user intent capture. Therefore, emphasizing implicit relation chain modeling and learning sequential correlations and dependencies among relations is novel and necessary for the recommendation task.

Recognizing the above challenges, we investigate diverse behavior patterns and examine how diverse auxiliary relations influence the target relation within behavior patterns, through the multiplex graph representation learning framework.
To this end, this work introduces a novel \textbf{\underline{D}}ual-\textbf{\underline{C}}hannel \textbf{\underline{M}}ultiplex \textbf{\underline{G}}raph \textbf{\underline{N}}eural \textbf{\underline{N}}etwork (\model) for multi-behavior recommendation tasks. We first design an \textit{explicit behavior pattern representation learner} to learn potential user preferences by explicitly modeling multi-behavior interactions based on the multiplex user-item bipartite graph. This capability empowers our developed recommender system to effectively model and leverage intricate behavior patterns characterized by multiplex user-item interaction networks.
Second, we introduce contrastive learning to learn the influences of different auxiliary relations on the target relation. Meanwhile, we capture correlations and dependencies between diverse relations in the behavior pattern.
The former achieves agreement between relation-specific embeddings through the constructed relation-based contrastive learning loss, and the latter considers that the correlations and dependencies between relations are of a certain order in a behavior pattern. 
Thus our model goes about exploring the appropriate order of relations in behavior patterns by designing an \textit{implicit relation chain effect learner} and a \textit{relation chain-aware contrastive learning} module. 
In particular, the ethical personalized knowledge extracted from the user is then used as a complement to the contrastive learning, and fed into the weighting function to guide contrastive learning to reinforce the differentiation of effects from various auxiliary relations on the target relation. 
Extensive experiments on three real-world datasets demonstrate the clear superiority of our proposed model over state-of-the-art (SOTA) baselines.

Our core innovation is a dual-channel multiplex GNN that learns node representations through: (1) behavior patterns in multiplex interaction networks, and (2) relation chains capturing relational dependencies. \textbf{For the first challenge}, the explicit behavior pattern representation learner employs hierarchical aggregation to model both local and global behavior patterns. \textbf{For the second challenge}, the implicit relation chain effect learner combines relation chain-aware contrastive learning to capture auxiliary relation effects on the target relation and dependencies and correlations between diverse relations. Finally, a joint optimization mechanism effectively integrates both explicit and implicit channels.

This work contributes the following:

\begin{itemize}[leftmargin=*]
    \item We introduce a multi-behavior recommendation method to emphasize the importance of constructing and exploiting user behavior patterns and address the issue of the influences of diverse auxiliary relations on the target relation. 
    \item We present an implicit relation chain effect learner to explore the effect of the sequence of different relations on the target relation. We also develop the relation chain-aware contrastive learning to enhance the understanding of the effects, correlations, and dependencies of different relation chains on the target relation. 
    \item Extensive experiments conducted on three real-world datasets show the efficacy of our proposed framework. The experiment results demonstrate that \model achieves up to 11.84\% and 14.34\% performance improvement compared to SOTA baselines in $Recall@10$ and $NDCG@10$. 
\end{itemize}

%% file: Main/2_Related_Work.tex
\vspace{-2mm}
\section{Related Work}
Multi-behavior recommendation~\cite{guo2023compressed,ma2023research,zhu2024multi,lee2024mule,liu2024multi} utilizes multiplex relations in user-item interactions~\cite{gao2019learning,jin2020multi,ren2024comprehensive}, which has garnered increasing attention recently because of its superiority in mitigating data sparsity and enhancing recommendation performance~\cite{meng2023coarse,rendle2012bpr,xia2021multi}.

Previous multi-behavior recommendation methods primarily rely on traditional recommendation techniques. A straightforward approach involves adapting a traditional matrix factorization technique designed for a single matrix to handle multiple matrices~\cite{singh2008relational,tang2016empirical}. 
For recommender systems such as HCCF~\cite{xia2022hypergraph} and LightGCL~\cite{cai2023lightgcl} that focus on only a single relation between the user and item, interactive information about other relations is discarded.

In contrast, another research idea is to treat multiplex relations as auxiliary and target relations, with novel sampling strategies developed for the training samples enrichment accordingly. Loni et al.~\cite{loni2016bayesian} introduce a method where different preference levels are assigned to multiple relations. They extend the BPR criterion~\cite{rendle2012bpr} by introducing a novel negative sampling strategy that includes sampling negative items from various relations. Gu et al. \cite{gu2022self} devise a self-supervised task aimed at distinguishing the importance of different relations to highlight distinctions between embeddings associated with these relations. They complement this approach with a star-contrastive learning paradigm designed to capture shared features between target and auxiliary relations. Guo et al.~\cite{guo2023compressed} leverage item-item similarity to generate samples across multiple auxiliary behaviors. In contrast, Huang et al.~\cite{xia2022hypergraph} on the other hand strengthens the information from collaborative filtering by using GCN.

With the rapid advancement of GNNs or graph convolutional networks (GCNs), they have been naturally integrated into the realm of recommender systems. Recent research efforts have also explored developing multi-behavior recommendation models based on these advancements. Typical GNN-based models initially learn user and item embeddings in each relation through a tailored network. Subsequently, they aggregate embeddings learned from various relations to predict the target relation~\cite{gao2019learning,xia2021graph,gao2019neural}. The difference is that different approaches will design different convolutional networks \cite{yu2022multiplex} and attention mechanisms \cite{peng2023attention}. Zhang et al. \cite{zhang2020multiplex} employ Multiplex Graph Neural Networks (MGNNs) to solve the problem of multi-behavior recommendation by performing the link prediction task. MGNN leverages the structures of multiplex networks and graph representation learning techniques to learn behavior-specific and shared embeddings for both users and items, enabling the modeling of collective effects from multiplex behaviors. Different from methods that aggregate embeddings from different relations to make the target relation predictions, NTMR~\cite{gao2019learning} develops a neural network model integrated with a multi-task learning framework to get intricate multi-type user-item interactions. The model considers cascading relations among diverse behavior categories. (\eg, users have to click before they can buy).

For GCN-based models, the typical approach involves constructing a unified U-I graph that incorporates all relations. Subsequently, GCN operations are applied on this graph to learn embeddings for users~\cite{jin2020multi,cheng2023multi,xia2021graph,xuan2023knowledge,yan2023mb}. MBGCN~\cite{jin2020multi} learns relation contributions on a unified U-I graph and in the I-I (item-item) interaction graph models relation semantics. The final prediction combines prediction scores derived from both relation contributions and relation semantics. MB-CGCN~\cite{cheng2023multi} is a multi-behavior recommendation model employing a cascading graph convolutional network. This model leverages behavior dependencies in embedding learning by using embeddings learned from one relation, after the feature transformation, as input features for the embedding learning of the next relation. MBGMN~\cite{xia2021graph} is a multi-behavior recommendation model that utilizes graph meta-networks, and it introduces a meta-learning framework that integrates multi-behavior pattern modeling, enabling the learning of user-item interactions to uncover topologically relevant relation representations. This approach automatically extracts interaction diversity and behavior heterogeneity to enhance recommendation effectiveness. Xuan et al. \cite{xuan2023knowledge} introduce the Knowledge-enhanced Multi-behavioral Contrastive Learning Recommendation (KMCLR), which incorporates a multi-behavior learning module for extracting personalized relation information about users for enhancing user embeddings. It also includes a knowledge enhancement module aimed at deriving robust perceptual representations of item knowledge. Additionally, an optimization phase is employed to model both coarse-grained similarities and fine-grained differences among users' multiplex relations. Except for NMTR~\cite{gao2019learning} and MB-CGCN~\cite{cheng2023multi}, Yan et al. \cite{yan2023cascading} introduce the CRGCN model, which addresses this limitation by employing a cascaded GCN structure for multi-task learning. However, because of its residual design, CRGCN is constrained to employing a single layer of GCNs for auxiliary behavior relations.

%% file: Main/3_Preliminary.tex
\vspace{-2mm}
\section{Preliminary}
\begin{figure*}
\label{framework}
    \begin{center}
    \includegraphics[width=1\textwidth]{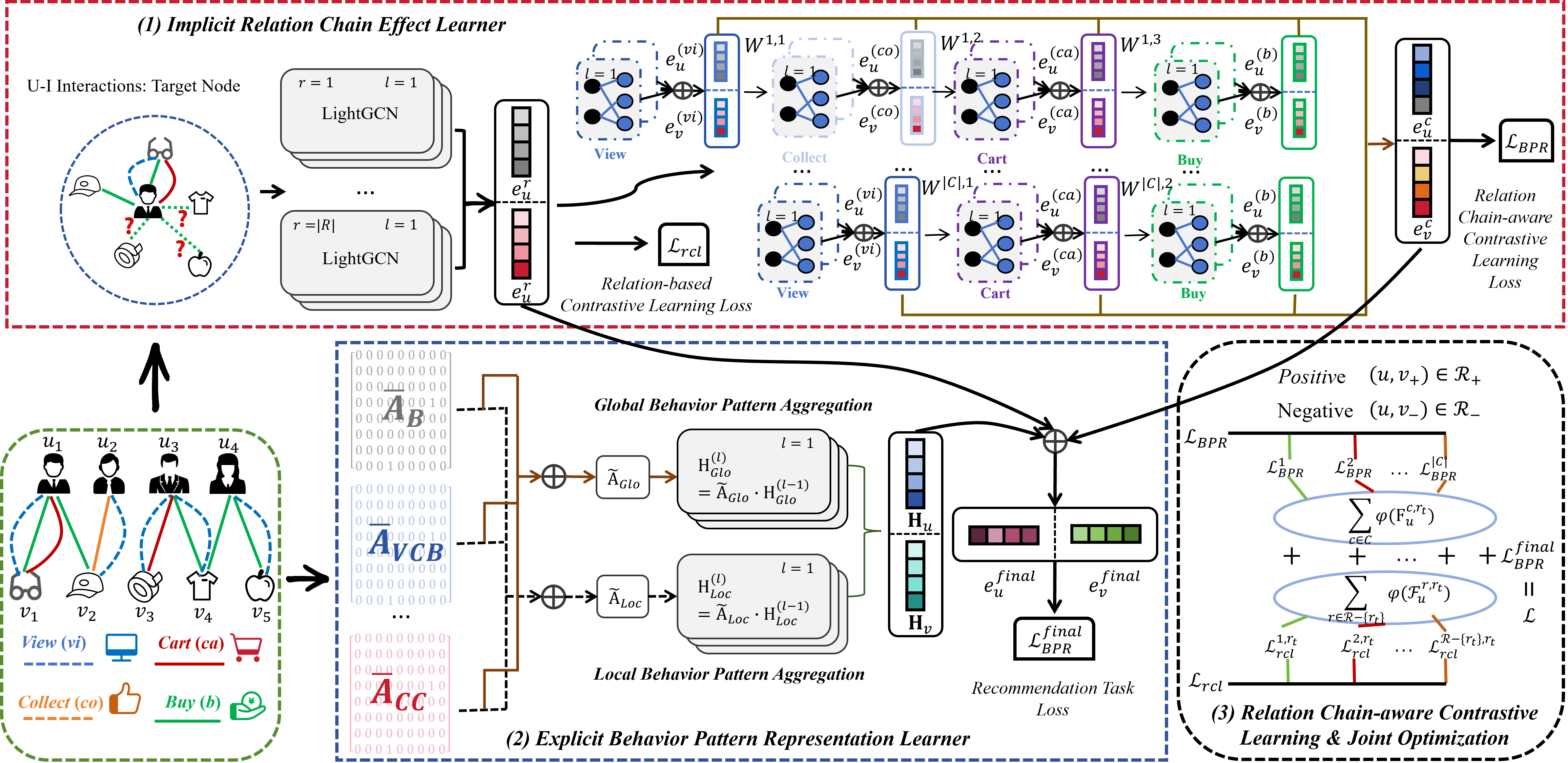}
    \caption{The overview of the proposed DCMGNN. $\Bar{A}_B$, $\Bar{A}_{VCB}$ and $\Bar{A}_{CC}$ are the BBP matrices containing corresponding relations, where $B$ denotes `Buy', $VCB$ denotes `View', `Cart' and `Buy', and $CC$ denotes `Collect' and `Cart'.}
    \label{fig:framework} 
    \end{center}
    \vspace{-6mm}
\end{figure*}

Given the sets of users and items along with multiple user-item interactive relations, we model these relationships using a multiplex bipartite graph as follows. 

\vspace{-2mm}
\begin{Def}[Multiplex Bipartite Graph]\label{def1}
We use $\mathcal{G}=\{\mathcal{U}, \mathcal{V}, \mathbf{X}, \mathcal{E}\}$ to represent a multiplex bipartite graph, where $\mathcal{U}$ and $\mathcal{V}$ respectively represent the user and item sets, $\mathbf{X} \in \mathbb{R}^{(|\mathcal{U}|+|\mathcal{V}|)\times f}$ denotes node (include uses and items) attribute feature matrix, and $\mathcal{E}=\bigcup\limits_{i\in \mathcal{R}} \mathcal{E}_i$ is the collection of various interactive edges between users and items, where each edge belonging to a particular interaction type, where $\mathcal{R}$ is the set of all interaction types. 
\end{Def}

For instance, in typical e-commerce networks in Fig.~\ref{fig:framework}, the interaction type set $\mathcal{R}$ usually includes viewing, carting, purchasing, etc. To capture the impact of multiple interactive relations on user and item representation, following~\cite{fu2023multiplex}, we also adopt the concept of the following basic behavior pattern.
\vspace{-2mm}
\begin{Def}[Basic Behavior Pattern, or BBP~\cite{fu2023multiplex}]\label{def2}
    BBPs between users $\mathcal{U}$ and items $\mathcal{V}$ in multiplex bipartite graphs can be defined as $\mathcal{U}\xrightarrow{{[{r_1}]\& [{r_2}]\& \ldots \& [{r_{|\mathcal{R}|}}]}} \mathcal{V}$ that represents a unique interaction behavior pattern between users and items, where $[\cdot]$ denotes optional, $|\mathcal{R}|$ denotes the number of relation types, and there exists at least one interactive relation type $r_i$. 
\end{Def}

For instance, $\mathcal{U} \xrightarrow{view} \mathcal{V}$ and $\mathcal{U} \xrightarrow{buy} \mathcal{V}$ are two types of BBPs in e-commerce datasets, which respectively represent that there is only one relation of `View' or `Buy' between users and items. 
$\mathcal{U} \xrightarrow{view\ \&\ buy} \mathcal{V}$ and $\mathcal{U} \xrightarrow{view\ \&\ cart\ \&\ buy} \mathcal{V}$ are also two types of BBPs,  
and the former means `View' and `Buy' relations co-occurring between users and items, while the latter means that there are three simultaneous interactive relations: `View', `Cart', and `Buy'. As depicted in Fig.~\ref{fig:framework}, the interactive behaviors between $u_3$ and $i_4$, $u_4$ and $i_5$ belong to pattern $\mathcal{U} \xrightarrow{view\ \&\ buy} \mathcal{V}$, the interactive behavior between $u_3$ and $i_3$ belongs to pattern $\mathcal{U} \xrightarrow{view\ \&\ cart} \mathcal{V}$, the interactive behavior between $u_1$ and $i_1$ belongs to pattern $\mathcal{U} \xrightarrow{view\ \&\ collect\ \&\ buy} \mathcal{V}$, and no interaction belongs to pattern $\mathcal{U} \xrightarrow{view} \mathcal{V}$. 

To further capture the influences of various relations on the target relation (\eg, `Buy' relation) in the same behavior pattern, we next introduce the concept of a relation chain.
\vspace{-2mm}
\begin{Def}[Relation Chain]
\label{def3}
    Given a BBP that includes at least two interactive relations, a relation chain is defined as a sequential order of all relations included in the BBP. 
\end{Def} 

Notice that, in general, BBPs must include the target relation (\eg, `Buy' or `Like' relation), and the last relation in the relation chain has to be the target relation.  
The guidelines for defining the sequential order of relation chains are implemented in the order of general principles of user behavior, and each BBP only claims one relation chain order. For example, we define the sequential order of interactive relations as $\langle view (\xrightarrow{} collect) \xrightarrow{} cart\xrightarrow{} buy \rangle$ in common e-commerce networks, and $\langle neutral \xrightarrow{} tips\xrightarrow{} like \rangle$ in review networks. For typical pattern $\mathcal{U} \xrightarrow{view\ \&\ cart\ \&\ buy} \mathcal{V}$, a corresponding relation chain $\langle view\xrightarrow{} cart\xrightarrow{} buy \rangle$ is produced.

Finally, we formally define our studied problem in this research as follows. 

\begin{Pro*}[]
    Given a multiplex bipartite graph $\mathcal{G}=\{\mathcal{U}, \mathcal{V}, \mathbf{X},  \mathcal{E}\}$, the goal of our recommender system is to learn a predictive function that estimates the likelihood of user $u \in \mathcal{U}$ interacting with item $v \in \mathcal{V}$ under the target interactive behavior. 
\end{Pro*}

The frequently-used critical symbols are in Table~\ref{tab.notation}.
\input{Tables/Notations}

%% file: Tables/Notations.tex
\begin{table}[htbp]
    \begin{center}
    \caption{Summary of frequently-used critical symbols.}
    \label{tab.notation}
    \setlength{\tabcolsep}{0.8mm}{}	
    \begin{tabular}{c|c}
        \toprule
       Symbol &  Definition  \\
        \midrule
        $\mathcal{G}$ & the input multiplex bipartite graph \\
        $\mathcal{U}, \mathcal{V}$ & the set of users/items \\
        $\mathbf{X}$ & the node attribute feature matrix\\
        $\mathcal{E}$ & the collection of various interactive edges in graph $\mathcal{G}$  \\
        $\mathcal{R}$ & the set of all interaction types in graph $\mathcal{G}$ \\
        $u,v$ & a user/item \\
        $\mathbf{A}_r$ & the adjacency matrix for the $r$-th relation of $\mathcal{G}$ \\
        $\Bar{\mathbf{A}}_{\zeta}$ & the adjacency matrix of the BBP type $\zeta$ \\
        $a_{\zeta},b_{\zeta}$ & the learnable weights for the BBP type $\zeta$ \\
        $\widetilde {\mathbf{A}}_{loc}$ & the local aggregation adjacency matrix \\
        $\mathbf{Q}$ & the matrix of global behavior pattern \\
        $\widetilde {\mathbf{A}}_{glo}$ & the similarity matrix of global behavior pattern  \\
        $\mathbf{H}_{loc}, \mathbf{H}_{glo}$ & the local and global hidden node representations \\
        $d$ & the embedding dimension \\ 
        $N, \mathbb{N}$ & the number of nodes/BBPs \\
        ${\mathcal{N}_u^r}, \mathcal{N}_v^r$ & the neighbor set of user/item for the $r$-th relation \\
        $e_{u/v}^{[r,l]}$ & user/item embedding for the $r$-th relation at $l$-th layer \\
        $e_{u/v}^{[r]}$ & user/item relation-specific embedding for the $r$-th relation \\
        $e_{u/v}^{[i,j]}$ & user/item embedding in $j$-th relation of the $i$-th relation chain\\
        $e_{u/v}^c$ & user/item embedding of relation chains\\
        $\mathbf{W}_{u/v}^{i,j}$ & the trans matrix of user/item in relation chains \\
        $C, C_i$ & the set of relation chains, the $i$-th relation chain \\
        $\mathbf{F}_u^{c,r_t}$ & the relation chain-aware knowledge  \\
        $\mathcal{F}_u^{r,r_t}$ & the relation-aware knowledge between embeddings  \\
        \bottomrule
    \end{tabular}
    \end{center}
    \vspace{-6mm}
\end{table}

%% file: Main/4_Method.tex
\vspace{-2mm}
\section{Methodology}\label{method}
In this section, we introduce the Dual-Channel Multiplex Graph Neural Network (\model) for multi-behavior recommendation, which contains three key components: \textit{(1) Explicit Behavior Pattern Representation Learner}, \textit{(2) Implicit Relation Chain Effect Learner}, and \textit{(3) Relation Chain-aware Contrastive Learning}. 
Underneath these, there are seven sub-modules, including \textit{(1.1) BBP Constructor}, \textit{(1.2) Local Behavior Pattern Aggregation}, \textit{(1.3) Global Behavior Pattern Aggregation}, \textit{(2.1) Multiplex Relation Embedding Aggregation}, \textit{(2.2) Relation Chain Representation Learning}, \textit{(3.1) Relation-based Contrastive Learning}, and \textit{(3.2) Relation Chain-based Contrastive Learning}. Specifically, \textit{BBP constructor} explicitly models multiplex network structure, \textit{local and global behavior pattern aggregations} effectively learn multi-relational representations of nodes. This addresses the first limitation. \textit{Multiplex relation embedding aggregation} and \textit{relation chain representation learning} capture the effect of auxiliary relations on the target relation. In addition, the proposed \textit{relation-based and relation chain-based contrastive learning} learn the correlations and dependencies between relations to further strengthen the influences of auxiliary relations to the target relation. This effectively addresses the second limitation. Fig.~\ref{fig:framework} illustrates the entire architecture of the proposed \model.

\vspace{-4mm}
\subsection{Explicit Behavior Pattern Representation Learner} 
There exist multiple interactive relations between users and items in multiplex bipartite graphs, and to better utilize these relations, we use the BBP, which contains a series of user-item interactions that represent the complete shopping behavior of a user for an item. Inspired by BPHGNN~\cite{fu2023multiplex}, a model for multiplex graph representation learning, we design an explicit behavior pattern representation learner, which explicitly captures the user-item interactive behavior into user/item representations. Three key components are included: \textit{BBP constructor}, \textit{local behavior pattern aggregation}, and \textit{global behavior pattern aggregation}. 

\subsubsection{BBP Constructor}

BBP constructor extracts all BBP matrices directly from the multiplex bipartite graph. Initially, the multiplex heterogeneous network structure is disentangled based on the types of edges (relations). We set $\{{\mathbf{A}}_r|r = 1,2,\ldots,|\mathcal{R}|\}$ as the adjacency matrix of the divided subgraphs. 
Next, each $\mathbf{A}_r$ and its corresponding logical variable (\ie, 0 or 1) undergoes an XNOR operation to get $|\mathcal{R}|$ intermediate matrices $\{ {\overleftrightarrow{\mathbf{A}}_r}|r = 1,2,\ldots,|\mathcal{R}|\}$. The logical variable is assigned the value of 1 to indicate preservation of the relation denoted by $\mathbf{A}_r$ in the BBP, and 0 otherwise. Finally, a per-position AND operation is performed on the intermediate matrices to obtain the BBP matrix. By traversing all logical variable combinations, we can obtain all BBP matrices $\{ {\overline{\mathbf{A}}_i} |i = 1,2,\ldots, 2^{|\mathcal{R}|}-1 \}$, and $\mathbf{A}_r,\overleftrightarrow{\mathbf{A}}_r,\overline{\mathbf{A}}_i \in \mathbb{R}^{N \times N}$, where $N = |\mathcal{U}| + |\mathcal{V}|$ denotes the all node number. 

\subsubsection{Local Behavior Pattern Aggregation}

Local behavior patterns are deep aggregations of BBPs, aiming at capturing the multiplex structure among nodes through the lens of local nodes, to mine the interactions among different relations in the behavior patterns. To achieve this goal, local behavior pattern aggregation incorporates an attention mechanism that assigns varying importance to each BBP through a set of learnable attention weights $a_i$, as shown below:
\begin{equation}
    {\widetilde {\mathbf{A}}_{loc}} = \sum\limits_{i = 1}^\mathbb{N} {{a_i}{\overline{\mathbf{A}}}_i},
\end{equation}
where $\mathbb{N}$ represents the number of BBPs, and \{$a_i$\} are randomly initialized according to a normal distribution, thus their values are also between 0 and 1.

Based on previous work, we use a simplified LightGCN~\cite{he2020lightgcn} to obtain representations, \ie, without linear activation and feature transformations: 
\begin{equation}
    \mathbf{H}_{loc}^{(l)} = {\widetilde {\mathbf{A}}_{loc}} \cdot \mathbf{H}_{loc}^{(l - 1)}.
\end{equation}

The propagation layer number determines the depth of the local behavior patterns. Eventually, the local node representation ${\mathbf{H}}_{loc} \in \mathbb{R}^{N \times d}$ is obtained by integrating the outputs from all layers to encompass the diverse interaction information within the behavior patterns across different depths as follows:
\begin{equation}
    \mathbf{H}_{loc} = \frac{1}{L}\sum\limits_{i = 1}^L \mathbf{H}_{loc}^{(i)}. 
\end{equation}

\subsubsection{Global Behavior Pattern Aggregation}
Global behavior patterns aggregate features across nodes from a global view, focusing on the similarity of BBPs among nodes. Users who exhibit similar behaviors tend to have similar purchase preferences. To be specific, we initially obtain a matrix to represent the global behavior patterns of users based on the acquired BBPs, adding rows to obtain column vectors for each BBP matrix, and each column vector means the frequency of BBPs across all users relative to each item. Given that various behavior patterns contribute unequally to similarity, we set a series of learnable weights $\{b_i\}$ to connect these column vectors for obtaining the global behavior pattern matrix $\mathbf{Q} \in \mathbb{R}^{N \times \mathbb{N}}$:
\begin{equation}
    {{\mathbf{Q}}_{\zeta(i)}} = \sum\limits_{q = 1}^{|\mathcal{V}|} {{{\overline {\mathbf{A}} }_{\zeta(i,q)}}}, 
\end{equation}

\begin{equation}
    {\mathbf{Q}} = \oplus_{i=1}^{\mathbb{N}} {b_i} \cdot {{\mathbf{Q}}_i}  = ( \oplus_{i=1}^{\mathbb{N}} {{\mathbf{Q}}_i}) \cdot {{\mathbf{\Lambda}}_b}, 
\end{equation}
where ${\mathbf{Q}}_{\zeta} \in \mathbb{R}^{N \times 1}$ represents the column vector which is associated with the $\zeta$-th BBP, $\oplus$ is the concatenation operation, \{$b_i\}$ are randomly initialized according to a normal distribution as \{$a_i$\}, and ${\mathbf{\Lambda}}_b = diag(b_1, b_2, \ldots, b_\mathbb{N})$ represents the learnable diagonal matrix.

Next, we obtain the similarity matrix $\widetilde {\rm A}_{glo}$ of global behavior pattern through transposition and normalization as follows:
\begin{equation}
    {\widetilde {\mathbf{A}}_{glo}} = norm({\mathbf{Q}} \cdot {{\mathbf{Q}}^\mathsf{T}}) \in {\mathbb{R}^{N \times N}}.
\end{equation}

We take the global behavior pattern similarity matrix as input into LightGCN for information aggregation to obtain node representations:
\begin{equation}
    {\mathbf{H}}_{glo}^{(l)} = {\widetilde {\mathbf{A}}_{glo}} \cdot {\mathbf{H}}_{glo}^{(l - 1)},
\end{equation}
and we obtain the global node representation ${\mathbf{H}}_{glo}$ by taking the output ${\mathbf{H}}_{glo}^{(l)} \in \mathbb{R}^{N \times d}$. 

We filter user embedding and item embedding using $\mathbf{H}_{loc}(u)$ and $\mathbf{H}_{loc}(v)$ respectively and then obtain the Explicit Behavior Pattern (EBP) embeddings $\mathbf{H}_{u/v}\in\mathbb{R}^{d}$ of user/item through the average pooling as:
\begin{equation}
    \mathbf{H}_{u/v} = \frac{1}{2}\left( {\mathbf{H}_{loc}}(u/v) + {\mathbf{H}_{glo}}(u/v) \right).
\end{equation}

To be specific, local behavior pattern aggregation represents the deep aggregation of BBPs. It is a composite pattern formed by multiple BBPs. Local behavior pattern aggregation performs feature aggregation from a local perspective by distinguishing the importance of different BBPs. Global behavior patterns aggregate features across nodes from a global view, focusing on the similarity of BBPs among these nodes.

\vspace{-4mm}
\subsection{Implicit Relation Chain Effect Learner}
In this section, we are to learn the impact of implicit relation chains within behavior patterns on node representation. First we learn multi-relational embeddings of users and items through LightGCN and then obtain node relation-chain embeddings by capturing the effect of relation chains corresponding to BBPs.

\subsubsection{Multiplex Relation Embedding Aggregation}

In this section, we use LightGCN~\cite{he2020lightgcn} as an information dissemination mechanism to learn the node embeddings from different relations. For the $r$-th relation in the multiplex bipartite graph, LightGCN utilizes user-item interactions to propagate embeddings as follows:
\begin{equation}
    e_u^{[r,l]} = \sum\limits_{v \in {\mathcal{N}_u^r}} {\frac{1}{{\sqrt {|{\mathcal{N}_u^r}| \cdot |{\mathcal{N}_v^r}|} }}e_v^{[r,l-1]}},
\end{equation}
\begin{equation}
    e_v^{[r,l]} = \sum\limits_{u \in {\mathcal{N}_v^r}} {\frac{1}{{\sqrt {|{\mathcal{N}_v^r}| \cdot |{\mathcal{N}_u^r}|}}}e_u^{[r,l-1]}},
\end{equation}
where $\mathcal{N}_u^r$ and $\mathcal{N}_v^r$ respectively denote the users' and items' neighbor sets under the $r$-th relation, $e_u^{[r,l]}$/$e_v^{[r,l]}$ represent the user/item embedding under the $r$-th relation throughout the propagation of $l$ LightGCN layers. After that, we simply aggregate the embeddings of user/item $\{ e_{u/v}^{[r,l]}\} _{l = 0}^L$ to get relation-specific embeddings: 
\begin{equation}
    e_{u/v}^{[r]} = \sum\limits_{l=0}^L{e_{u/v}^{[r,l]}}.
\end{equation}

Finally, we obtain the multi-relation embeddings of users and items as follows:
\begin{equation}
    e_{u/v}^r = \sum\limits_{r \in \mathcal{R}} {e_{u/v}^{[r]}}. 
\end{equation}

\subsubsection{Relation Chain Representation Learning}

According to Definition~\ref{def3} (\ie, Relation Chain), a relation chain is generated for each basic behavior pattern (BBP) containing the target relation (\ie, buy relation) and two or more relations following the predefined sequential order. Let ${\mathbf{W}}_u^{i,j}$ and ${\mathbf{W}}_v^{i,j}$ be the learnable transformation matrix parameters of users and items from the $j$-th relation to the $(j+1)$-th relation in the $i$-th relation chain, and the transformation is performed as:
\begin{equation}
    e_{u/v}^{[i,j+1]} = {\mathbf{W}}_{u/v}^{i,j}e_{u/v}^{[i,j]}, 
\end{equation}
where $e_{u/v}^{[i,j]}$ is the embedding of user/item in $j$-th relation. When the transformation process is over, we will finally get the relation-chain embeddings $e_u^c$ and $e_v^c$ of user $u$ and item $v$ across all relation chains as follows:
\begin{equation}
    e_{u/v}^c = \sum\limits_{i = 1}^{|C|} {\sum\limits_{j = 1}^{|C_i|} {e_{u/v}^{[i,j]}}},
\end{equation}
where $C=\{C_1,C_2,\dots,C_m\}$ represents the set of relation chains, and $m$ denotes the number of relation chains. According to the experiments, it is proved that the feature transformation process among the relation chains is simple and effective, and the features can be extracted effectively. 

According to the learned embeddings $\mathbf{H}_{u/v}$, multi-relation-based embeddings $e_{u/v}^r$, and relation-chain embeddings $e_{u/v}^c$, we can get the final embeddings of user/item as follows:
\begin{equation}
    e_{u/v}^{final} = Mean\left( \mathbf{H}_{u/v}, e_{u/v}^r, e_{u/v}^c \right).
\end{equation}

\vspace{-4mm}
\subsection{Relation Chain-aware Contrastive Learning}\label{rc-ac}
In e-commerce networks, different users have different interaction preferences in their behavior patterns toward items. In multiplex bipartite graphs, multiple behavior patterns lead to different item-relational interactions for different users, so it is critical to effectively model the dependencies and correlations between different types of relations in behavior patterns. 

Before that, it is also important to consider the multiple user-item interactive relations contained in different behavior patterns, so we first set up a relation-based contrastive learning module for distinguishing between the target relation and other different auxiliary relations. Specifically, we define different relations involving the same user as positive samples, while different users are viewed as negative samples. For user-item interactions, the intuition is represented by different interactive relations, so setting up such a comparison learning module helps us to capture the correlations between auxiliary relations (\eg, `View', `Cart') and the target (`Buy') relation.

To achieve this goal, we introduce a relation chain-aware contrastive learning framework for integrating explicit weighting functions for contrastive losses of different relations in multiple BBPs. The relation chain-aware contrastive learning is divided into two phases: First, we use a relation-aware encoder to represent different relation chains to extract the multi-relational features of U-I interactions, reflecting user shopping preferences under different relation chains. Second, the knowledge extracted from the relation-aware encoder serves as an input to a relation chain-aware network that generates customized contrastive loss weights for the relation chain dependency modeling. 

\subsubsection{Relation-based Contrastive Learning}

Following existing methods, we use the InfoNCE loss in the relation-based contrastive learning module for measuring the difference between relation-specific embeddings. We calculate the relation-based contrastive loss:
\begin{equation}
    \mathcal{L}_{rcl}^{r',r} = \sum\limits_{u \in \mathcal{U}} { - \log \frac{{\exp (\phi (e_u^{(r)},e_u^{(r')})/\tau )}}{{\sum\nolimits_{u' \in {\mathcal{U}-u}} {\exp (\phi (e_u^{(r)},e_{u'}^{(r')})/\tau )}}}}.
\end{equation}

Here $\phi(\cdot)$ is the similarity function (such as the inner-product operation) between contrastive embeddings, and $\tau$ is the hyperparameter. Generally, we maximize the contrastive loss to make the different relations more distinguishable from each other than before and use this to measure the variability between different users. 

\subsubsection{Relation Chain-based Contrastive Learning}
In this module, we first extract the relation-aware knowledge to preserve relation dependencies and correlations. Inspired by the feature extraction mechanism in CML~\cite{wei2022contrastive}, we set two kinds of relation-aware encoder within two types of aggregation techniques based on the learned users' relation embeddings, relation-chain embeddings, and final embeddings as follows:
\begin{equation}
    \mathbf{F}_u^{c,r_t} = (dup(\sum\limits_{r \in c}{\mathcal{L}_{rcl}^{r, r_t}}) \cdot \mu) \oplus e_u^{c} \oplus {e_u^{final}},
\end{equation}
\begin{equation}
    \mathcal{F}_u^{r,r_t} ={\mathcal{L}_{rcl}^{r,r_t}} \cdot (e_u^{(r)} \oplus {e_u^{final}}),
\end{equation}
where $r_t$ denotes the target relation, and $\mathbf{F}_u^{c, r_t}$ represents the dependencies that retain user preference information between all relations in the user relation chain and the target relation. The $dup(\cdot)$ refers to the duplicate function generating a vector with the same embedding dimension as the original embedding. $\oplus$ represents the embedding concatenation function, $\mu$ represents the scale factor of the enlargement value. $\mathcal{F}_u^{r, r_t}$ cooperatively represents the relation-aware knowledge between relation-specific embedding and user embedding. 
According to this, relation dependencies and correlations in the relation chain and relation-specific embeddings can be preserved.

Next, we transform the preserved information into relation chain-aware contrastive weights. Let $\epsilon(\cdot)$ represent the transformation function, and we can obtain the relation dependencies and correlations $\epsilon ({\mathbf{F}}_u^{c,r_t})$ as follows:
\begin{equation}
    \epsilon ({\mathbf{F}}_u^{c,r_t}) = LeakyReLU({\mathbf{F}}_u^{c,r_t} \cdot {{\mathbf{W}}^\epsilon } + {b^\epsilon }),
\end{equation}
where ${\mathbf{W}}^\epsilon$ and $b^\epsilon$ represent the projection and bias items, and we can obtain $\epsilon(\mathcal{F}_u^{r,r_t})$ by obeying the same way. 

\vspace{-4mm}
\subsection{Joint Optimization}
In this section, we outline the objective of \model. We leverage the BPR loss, which can be formally defined, to learn corresponding parameters in the model inference as follows:
\begin{equation}
    \mathcal{L}_{BPR}^c = \sum\limits_{(u,{v_ + },{v_ - }) \in {\mathcal{T}_c}} { - \ln (Sigmoid(\hat{y}_{u,{v_ + }}^c - \hat{y}_{u,{v_ - }}^c))}  + \lambda ||\Theta |{|^2}
\end{equation}
where $\hat{y}_{u,{v_+}}^c = {e_u^c}^\mathsf{T}{e_{v_+}^c}$, $\mathcal{T}_c$ is the training samples of $c$-th relation chain, \ie, $\mathcal{T}_c=\{(u,{v_ + },{v_ - })|(u,{v_ + }) \in \mathcal{O}_+, (u,{v_ - }) \in \mathcal{O}_-\}$, $\mathcal{O}_+$ and $\mathcal{O}_-$ respectively denote the observed existent interaction and unobserved interaction in the corresponding behavior pattern. $\Theta$ represents the learnable parameter, we use $L_2$ regularization to prevent over-fitting and $\lambda$ to control it. 

\begin{equation}
    {\mathcal{L}_{BPR}} = \sum\limits_{c \in C} \epsilon ({\mathbf{F}}_u^{c,r_t})  \cdot \mathcal{L}_{BPR}^c, 
\end{equation}
\begin{equation}
    {\mathcal{L}_{rcl}} = \sum\limits_{r \in \mathcal{R} - \{ {r_t}\} } \epsilon (\mathcal{F}_u^{r,r_t})  \cdot \mathcal{L}_{rcl}^{r,r_t},
\end{equation}
where $r_t$ is the target relation and $C$ is the set of all relation chains. 
\begin{equation}
    \mathcal{L}_{BPR}^{final} = \sum\limits_{(u,{v_ + },{v_ - }) \in \mathcal{O}} { - \ln (Sigmoid(\hat{y}_{u,{v_ + }} - \hat{y}_{u,{v_ - }}))}  + \lambda ||\Theta |{|^2}.
\end{equation}

Here $\hat{y}_{u,{v_ +}} = {e_u^{final}}^\mathsf{T}{e_{v_+}^{final}}$, $\mathcal{O}$ denotes the set of all positive and negative samples. The final loss of the proposed \model is as follows:
\begin{equation}\label{eq:final_loss}
    \mathcal{L} = \mathcal{L}_{BPR}^{final} + \lambda_{1}{\mathcal{L}_{rcl}} + \lambda_{2}\mathcal{L}_{BPR}
\end{equation}

\vspace{-4mm}
\subsection{Time and Space Complexity Analysis}\label{complexity}

Our \model contains three main components: explicit behavior pattern representation learner, implicit relation chain effect learner, and relation chain-aware contrastive learning.

\textit{1) Time Complexity Analysis:}
For the explicit behavior pattern representation learner, the time complexity is $O(\mathbb{N}N^2 + Nd(NL + d(L-1) +1))$, where $N$ and $\mathbb{N}$ are the number of nodes and BBPs, and $L$ is the aggregation layer count. For the implicit relation chain effect learner, the time complexity is $O(NdL+md^2+md)$, where $m$ denotes the relation chain count. The relation chain-aware contrastive learning has a time complexity of $O(Nd^2 + Nd)$. Therefore, the entire time complexity for DCMGNN is approximately $O((\mathbb{N}+dL)N^2 + d^2LN + md^2)$. Additionally, multiplex network representation learning for explicit and implicit channels can also be accelerated in a parallel manner.

\textit{2) Space Complexity Analysis:}
For the explicit behavior pattern representation learner module, the input of the local behavior pattern aggregation includes diverse BBPs matrices, and the weight vector $a_{1:\mathbb{N}}$, thus the space complexity of local behavior pattern aggregation is $O(\mathbb{N}N^2 + \mathbb{N})$. Similarly, the input of the global behavior pattern aggregation contains the matrix $Q$, the adjacency matrix $\widetilde {\mathbf{A}}_{glo}$, and the weight vector $b_{1:\mathbb{N}}$, thus the space complexity of global behavior pattern aggregation is $O(N^2 + \mathbb{N}N + \mathbb{N})$. The total space complexity of the explicit behavior pattern representation learner is $O((\mathbb{N} + 1)N^2 + \mathbb{N}N + \mathbb{N})$. For the implicit relation chain effect learner module, the input of the multiplex relation embedding aggregation includes user/item embeddings of $|\mathcal{R}|$ relations after $L$ layers LightGCN message passing, thus the space complexity is $O(L|\mathcal{R}|Nd)$. Similarly, the space complexity of the relation chain representation learning is $O(m|\mathcal{R}|Nd)$. Therefore, the total space complexity of the implicit relation chain effect learner is $O((m + L)d|\mathcal{R}|N)$. For the relation chain-aware contrastive learning module, the space complexity of the relation-based contrastive learning is $O(N^2(|\mathcal{R}|-1)d)$, the space complexity of the relation chain-based contrastive learning is $O(N(|\mathcal{R}|-1)d + d^2 + d)$. Therefore, the total space complexity of the relation chain-aware contrastive learning is $O((|\mathcal{R}|-1)dN^2 + (|\mathcal{R}|-1)dN + d^2 + d)$. To sum up, the total space complexity of DCMGNN is approximately $O((|\mathcal{R}|d + \mathbb{N})N^2 + ((m + L)|\mathcal{R}|d + \mathbb{N})N + \mathbb{N} + d^2 + d)$.

%% file: Main/5_Experiments.tex
\vspace{-2mm}
\section{Experiments}
\subsection{Datasets and Evaluation Metrics}
\noindent
\textbf{Datasets}: Three publicly available real-world datasets are utilized in our experiments, \ie, Retail\_Rocket (Retail for short)~\cite{ren2023sslrec}, Tmall~\cite{cheng2023multi}, and Yelp~\cite{gu2022self}. Table~\ref{tab:datasets} summarizes the statistics, and we detail the dataset description as follows: 
\begin{itemize}[leftmargin=*]
    \item \textbf{Retail~\cite{ren2023sslrec}:} It is a benchmark dataset collected from the Retail\_rocket recommender system and includes user interactions page views (View), add-to-cart (Cart), and transactions (Buy). According to previous research on multi-behavior recommendations~\cite{xia2020multiplex,jin2020multi}, `Buy' is the target relation.
    \item \textbf{Tmall~\cite{cheng2023multi}:} This dataset includes three interactive relations, \ie, page viewing (View), add-to-cart (Cart), and purchasing (Buy), where the `Buy' is the target relation.
    \item \textbf{Yelp~\cite{gu2022self}:} This dataset is collected from Yelp, which contains four interactive relations, \ie, like (Like), tip (Tips), neutral (Neutral), and dislike (Dislike), where the `Like' is the target relation.
\end{itemize}
\input{Tables/Datasets}

\noindent
\textbf{Evaluation Metrics}: We assess our \model and baseline methods based on the top-$k$ recommended items in all experiments using the following metrics: the Recall (abbreviated as $R@5$, $R@10$, $R@20$, $R@40$) and the Normalized Discounted Cumulative Gain (NDCG) (abbreviated as $N@5$, $N@10$, $N@20$, $N@40$). 

\vspace{-4mm}
\subsection{Baselines}\label{sec.baseline}
We compare our \model with the following 22 baselines, which are categorized into two main groups: 

\textbf{\textit{Single-Behavior Recommendation Methods}}
\begin{itemize}[leftmargin=*]
    \item \textbf{BPR \cite{rendle2012bpr}:} This model is widely applied to matrix factorization, leveraging Bayesian personalized ranking as the optimization criterion.
    \item \textbf{LightGCN \cite{he2020lightgcn}:} It simplifies the GCN-based recommendation framework by removing nonlinear activation and feature transformation.
    \item \textbf{HCCF \cite{xia2022hypergraph}:} It introduces a hypergraph-enhanced cross-view contrastive learning to simultaneously capture local and global collaborative relations.
    \item \textbf{DCCF \cite{ren2023disentangled}:} This model disentangles intent via adaptive self-supervised augmentation and cross-view contrastive learning, extracting fine-grained factors and reducing noise.
    \item \textbf{AutoCF \cite{xia2023automated}:} It autonomously augments data, prioritizes generative self-supervised learning, and uses learnable augmentation methods to extract key signals.
    \item \textbf{LightGCL \cite{cai2023lightgcl}:} 
    It enhances the generality and robustness of CL-based recommendations using singular value decomposition and global collaborative relation modeling for structural refinement.
\end{itemize}

\textbf{\textit{Multi-Behavior Recommendation Methods}}
\subsubsection{\textbf{GNN-based methods}}
\begin{itemize}[leftmargin=*]
    \item \textbf{RGCN \cite{schlichtkrull2018modeling}:} It differentiates node relations by edge types and uses separate message propagation layers for each, enabling adaptation to multi-behavior recommendations.
    \item \textbf{MBGCN \cite{jin2020multi}:} It captures multi-behavior relations across the U-I interaction graph it constructs, incorporating high-order connectivity in its information propagation strategy.
    \item \textbf{MB-HGCN \cite{yan2023mb}:} It employs a hierarchical GCN to learn node embeddings by progressing from global-level coarse-grained representations to behavior-level finer-grained ones. 
    \item \textbf{MB-CGCN \cite{cheng2023multi}:} It uses cascading graph convolutional networks, where behavior embeddings are transformed for learning subsequent behaviors.
    \item \textbf{CRGCN \cite{yan2023cascading}:} It models multi-behavior data with a cascading GCN, passing features from earlier behaviors to later ones via a residual design, and optimizes the model using multi-task learning.
    \item \textbf{BPHGNN \cite{fu2023multiplex}:} It is a multiplex GNN method tailored for obtaining embeddings of nodes in multi-behavior recommendation tasks through specific behavior pattern modeling.
    \item \textbf{S-MBRec \cite{gu2022self}:} It leverages GCNs to learn node embeddings and a supervision task to differentiate behavior importance, enhancing embedding distinctions.
\end{itemize}
\subsubsection{\textbf{Multi-task/Meta Learning-based methods}}
\begin{itemize}[leftmargin=*]
    \item \textbf{NMTR \cite{gao2019learning}:} It integrates multi-task learning with neural collaborative filtering to examine multiple types of user interactive behaviors using predefined cascading relations.
    \item \textbf{IMGCF \cite{zhang2023alleviating}:} It tackles data imbalance in multi-behavior graph filtering using multi-task learning to enhance sparse behavior representation.
    \item \textbf{MBGMN \cite{xia2021graph}:} It enhances user-item interaction learning by capturing type-dependent behavior representations and interaction diversity for recommendations.
    \item \textbf{MBA \cite{xin2023improving}:} It learns user preferences from implicit feedback by maximizing likelihood and minimizing KL divergence between behavior models.
    \item \textbf{HPMR \cite{meng2023hierarchical}:} It features the Projection-based Transfer Network (PTN), which refines upstream behavior representations by modeling correlations between behaviors and enhances downstream task learning.
\end{itemize}
\subsubsection{\textbf{Contrastive Learning (CL)-based methods}}
\begin{itemize}[leftmargin=*]
    \item \textbf{HMG-CR \cite{yang2021hyper}:} It introduces the concept of hyper-metric paths or hyper-metric graphs to explicitly explore the dependencies between diverse user behaviors.
    \item \textbf{MMCLR \cite{wu2022multi}:} It includes three tasks: aligning user representations, bridging sequence and graph views, and modeling behavior differences.
    \item \textbf{CML \cite{wei2022contrastive}:} This framework uses multi-behavior comparative learning to extract transferable knowledge across behavior categories with a custom contrastive loss function.
    \item \textbf{KMCLR \cite{xuan2023knowledge}:} It enhances user embeddings with multi-behavior learning and improves item representations using a knowledge graph. 
\end{itemize}

\input{Tables/Baselines}
\input{Tables/Supply_Baselines}
\vspace{-4mm}
\subsection{Experimental Setting}
We designate the connected user-item node pairs as positive samples and consider other unconnected node pairs as negative ones for recommendation. We partition the positive samples into the training, validation, and test sets on the specified proportions of 70\%, 10\%, and 20\%. 

We randomly select an equal number of negative samples to add to both training and test sets. In our experiments, we optimize our models using the Adam optimizer. The embedding dimension $d$ and batch size are set from \{8, 16, 32, 64, 128, 256\} and 128 for all methods, while the learning rate varies within the range \{1e-4, 1e-3, 1e-2, 1e-1\}. In addition, we initialize the model parameters (specifically, the relation feature transformation matrix) by using the \textit{Xavier} initializer. The scale factor $\mu =
0.98$ of the enlargement value is selected from \{0.95, 0.96, 0.97, 0.98, 0.99, 1.00\}, and the temperature hyperparameter $\tau$ is selected from \{10, 5, 1, 0.5, 0.1, 0.05\} and set as 0.1. The number of LightGCN layers per relation varies within the range \{1,2,3,4\}. 
Unless otherwise stated, for multiple relations in a relation chain, we use solid three LightGCN layers. 

We utilize the provided source code of baselines and adhere to the parameter settings recommended in their respective papers to make sure that their methods can achieve results closely aligned with their reported claims. We have conducted each experiment ten times and report the average values. 

For Tmall and Retail datasets, they contain three relations: `View', `Cart', and `Buy', where `Buy' is the target relation, and the defined relation chain order is $\langle View\xrightarrow{} Cart\xrightarrow{} Buy \rangle$.  
Notice that, as stated in Definition 3, relation chains depend on BBPs. There will be as many relation chains as there are BBPs containing the `Buy' relation in each dataset. Note that the order of relations in all relation chains obeys $\langle View\xrightarrow{} Cart\xrightarrow{} Buy \rangle$. For Yelp dataset, the relation `Like' is defined as the target relation, and the relation chain order is $\langle Neutral \xrightarrow{} Tips \xrightarrow{} Like \rangle$. There exists behavior patterns such as $\mathcal{U} \xrightarrow{tips\ \&\ like} \mathcal{V}$ and $\mathcal{U} \xrightarrow{neutral\ \&\ tips\ \&\ likes} \mathcal{V}$ in review networks like Yelp.

\vspace{-4mm}
\subsection{Overall Performance}
We perform a performance comparison between \model and all baselines, with results shown in Table~\ref{tab:perform_compared} and~\ref{tab:supply_perform_compared}. The best results are highlighted in bold, and the second-best results are underlined. \textit{Improvement} denotes the improvement of our results compared with the second-best results.

As we can observe from the tables, our \model significantly outperforms all baselines in all metrics on the three datasets, with specific emphasis on the improvement relative to the SOTA methods. \model yields notable enhancements, achieving a 14.34\% improvement in terms of $N@10$ on Retail, a 11.84\% improvement in terms of $R@10$ on Tmall, and a 12.12\% improvement in terms of $N@10$ on Yelp. It is noteworthy that the well-performed top-3 baselines, CML, KMCLR, and MB-CGCN, have already demonstrated substantial improvements on these datasets, providing strong evidence of the effectiveness of our \model. 

For multi-behavior recommendation models, most of them have achieved better results than single-behavior recommendation models. These approaches aim to model the intricate user-item relations through graph structure learning and capture higher-order interactive information that is difficult to access by collaborative filtering, and also achieve refreshing results. In addition, some works based on MBGCN (\eg, MB-RGCN, MB-CGCN, MBGMN, and CRGCN), intend to obtain more U-I interactive information by improving the graph convolution mechanism. CML captures the dependency between various auxiliary relations and the target relation through the improvement of contrasting learning, and KMCLR supplements the missing node by introducing the knowledge graph attribute features. 

The superior performance of our method can be attributed to two primary aspects.
First, we model the multi-behavior recommendation problem as the representation learning of a multiplex bipartite graph and devise a dual-channel multiplex heterogeneous GNN. One channel leverages BBPs to explicitly learn node representations of the multiplex graph, and another channel uses relation chains to implicitly capture the influences of other auxiliary relations on the target relation. The final representations of nodes combine the learned node embeddings in terms of multiple factors.
Second, in contrastive learning, we not only consider the differences and alignments between various relations but also integrate the dependencies and correlations extracted from the relation chains into the loss function of contrastive learning. This enhances the contribution of auxiliary relations to the target relation. These designs allow us to capture the impact of multiple heterogeneous relations on node representation from different aspects, as well as the correlations and dependencies between these relations. 

Conclusively, the final experimental evaluation further validates the effectiveness, superiority, and novelty of \model in capturing the effect of explicit multi-relational behavior patterns and the influence of implicit relation chains on target relations simultaneously. 

\begin{figure*}
    \begin{center}
    \includegraphics[width=0.96\textwidth]{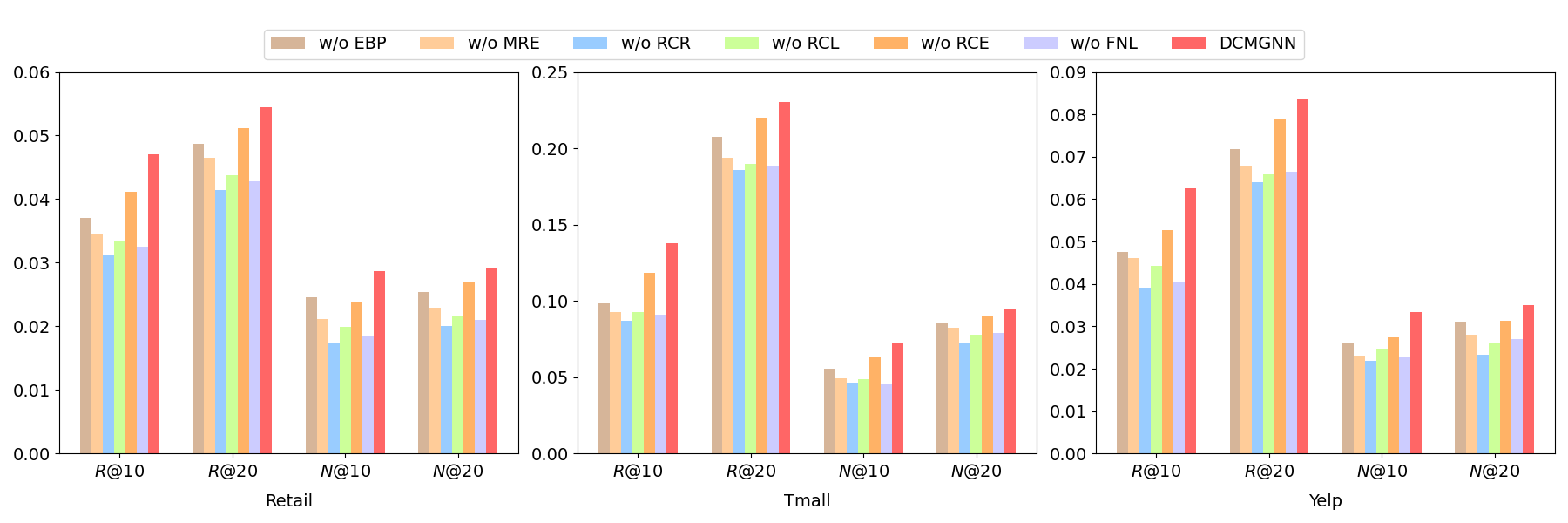}
    \caption{The comparison of DCMGNN and its variants.}
    \label{fig:ablation} 
    \end{center}
    \vspace{-6mm}
\end{figure*}

\begin{figure*}[h]
    \centering
    \captionsetup[subfigure]{font=footnotesize, labelfont=bf}
    \begin{subfigure}{0.32\textwidth}
        \includegraphics[width=\linewidth]{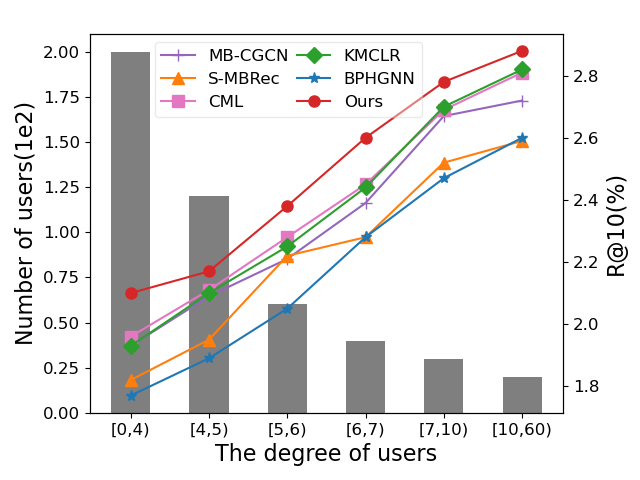}
        \caption{$R@10$ in Retail}
        \label{fig:sparity_Retail_R}
    \end{subfigure}
    \hspace{-2mm}
    \begin{subfigure}{0.32\textwidth}
        \includegraphics[width=\linewidth]{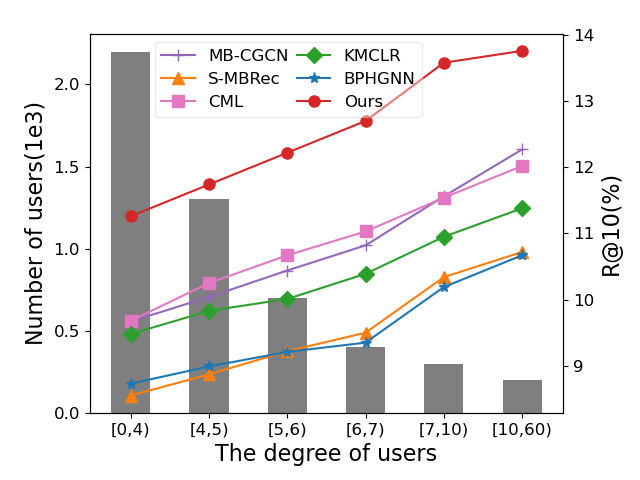}
        \caption{$R@10$ in Tmall}
        \label{fig:sparity_Tmall_R}
    \end{subfigure}
    \hspace{-2mm}
    \begin{subfigure}{0.32\textwidth}
        \includegraphics[width=\linewidth]{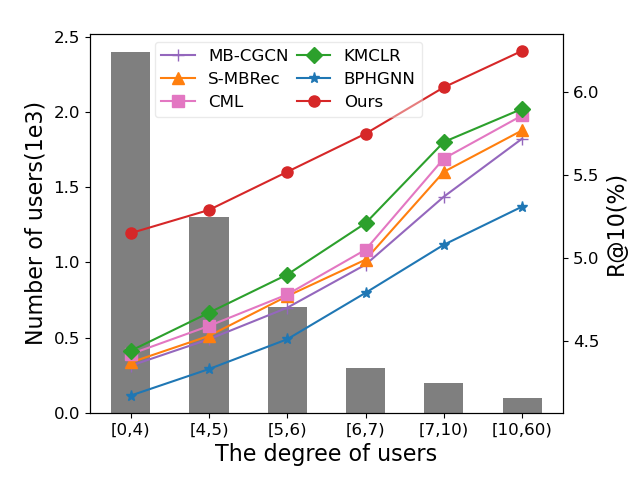}
        \caption{$R@10$ in Yelp}
        \label{fig:sparity_Yelp_R}
    \end{subfigure}
    \begin{subfigure}{0.32\textwidth}
        \includegraphics[width=\linewidth]{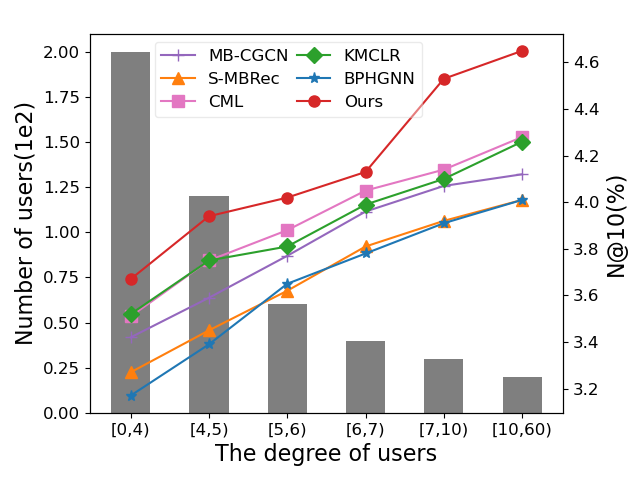}
        \caption{$N@10$ in Retail}
        \label{fig:sparity_Retail_N}
    \end{subfigure}
    \hspace{-2mm}
    \begin{subfigure}{0.32\textwidth}
        \includegraphics[width=\linewidth]{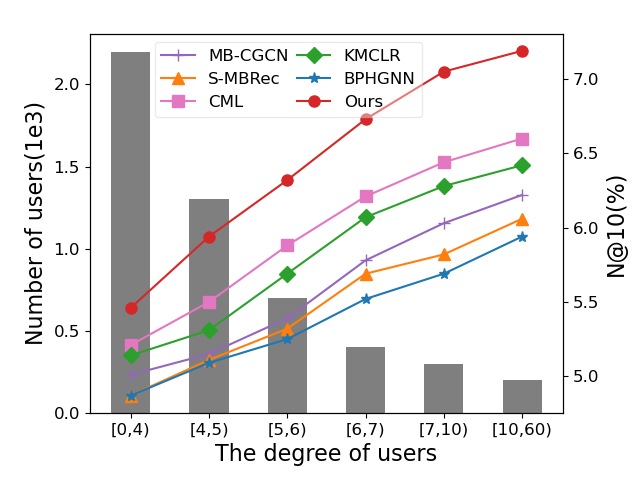}
        \caption{$N@10$ in Tmall}
        \label{fig:sparity_Tmall_N}
    \end{subfigure}
    \hspace{-2mm}
    \begin{subfigure}{0.32\textwidth}
        \includegraphics[width=\linewidth]{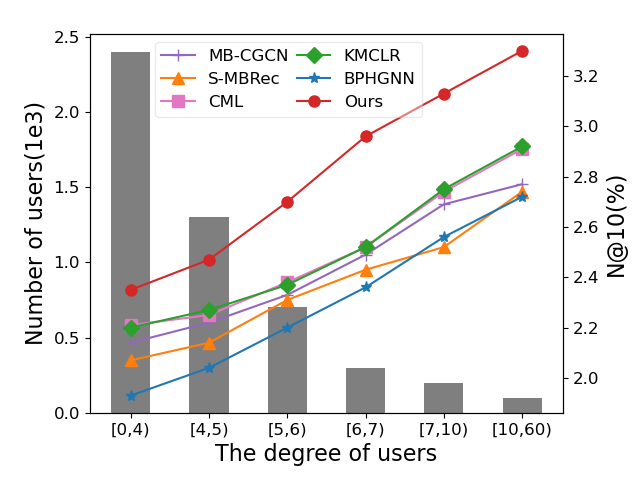}
        \caption{$N@10$ in Yelp}
        \label{fig:sparity_Yelp_N}
    \end{subfigure}
    \caption{Performance comparison \wrt various degrees of relation sparsity on different datasets.}
    \label{fig:sparsity_T}
    \vspace{-6mm}
\end{figure*}
\begin{figure}[t]
    \centering
    \captionsetup[subfigure]{font=footnotesize, labelfont=bf}
    \begin{subfigure}{0.23\textwidth}
        \includegraphics[width=\linewidth]{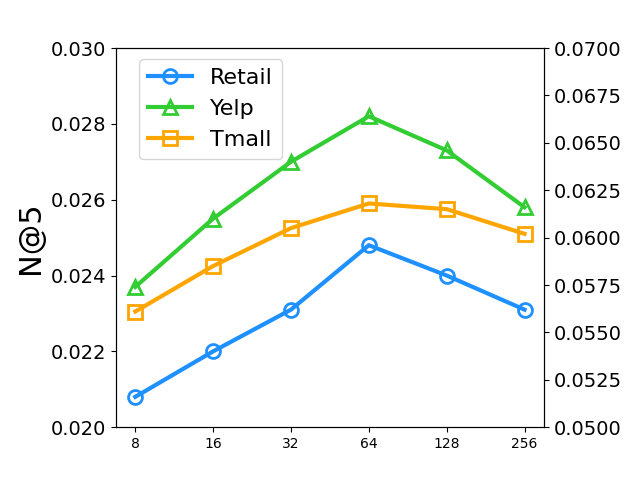}
        \caption{$N@5$ \wrt $d$}
        \label{fig:ndcg5_dim}
    \end{subfigure}
    \hspace{-2mm}
    \begin{subfigure}{0.23\textwidth}
        \includegraphics[width=\linewidth]{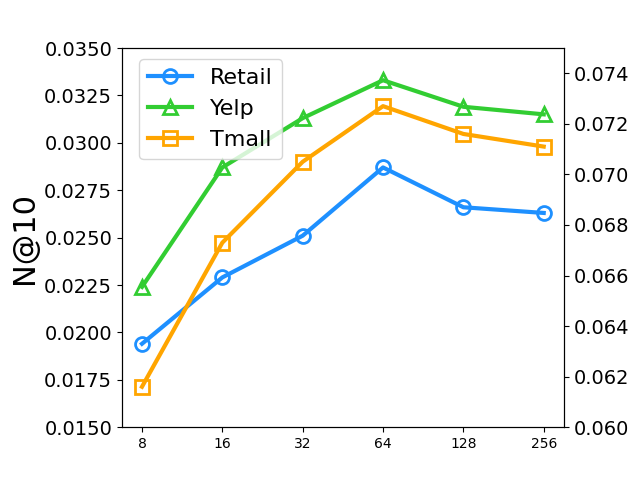}
        \caption{$N@10$ \wrt $d$}
        \label{fig:ndcg10_dim}
    \end{subfigure}
    \begin{subfigure}{0.23\textwidth}
        \includegraphics[width=\linewidth]{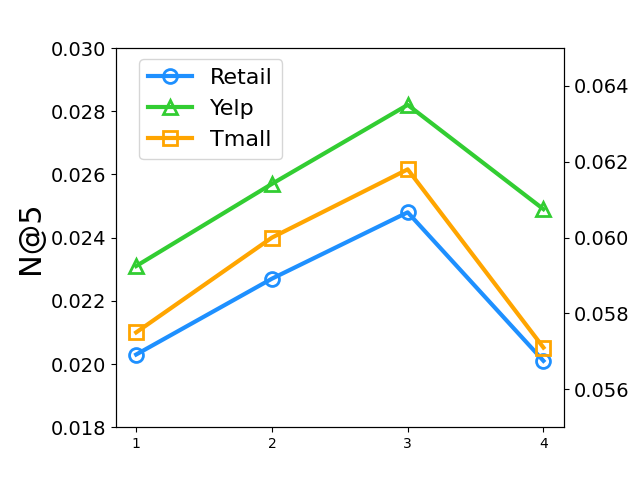}
        \caption{$N@5$ \wrt $L$}
        \label{fig:ndcg5_l}
    \end{subfigure}
    \hspace{-2mm}
    \begin{subfigure}{0.23\textwidth}
        \includegraphics[width=\linewidth]{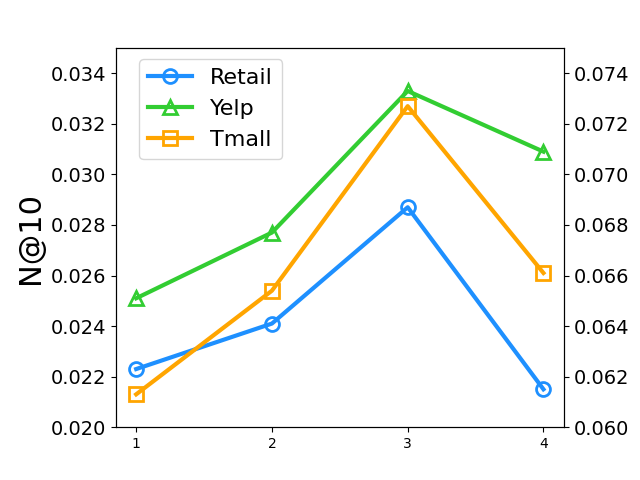}
        \caption{$N@10$ \wrt $L$}
        \label{fig:ndcg10_l}
    \end{subfigure}
    \caption{The impact of embedding dimension $d$.}
    \label{fig:para1}
    \vspace{-5mm}
\end{figure}
\begin{figure}[t]
    \centering
    \begin{subfigure}{0.23\textwidth}
        \includegraphics[width=\linewidth]{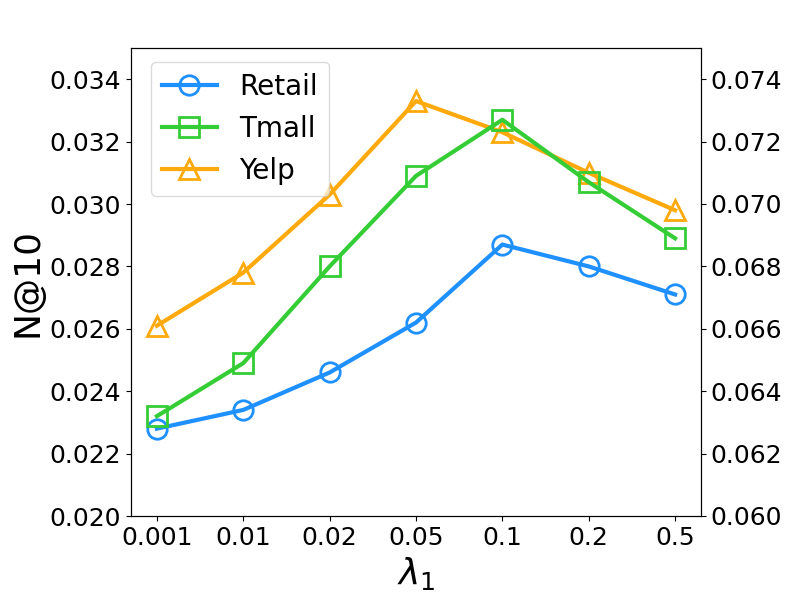}
        \label{fig:lambda_1}
    \end{subfigure}
    \hspace{-2mm}
    \begin{subfigure}{0.23\textwidth}
        \includegraphics[width=\linewidth]{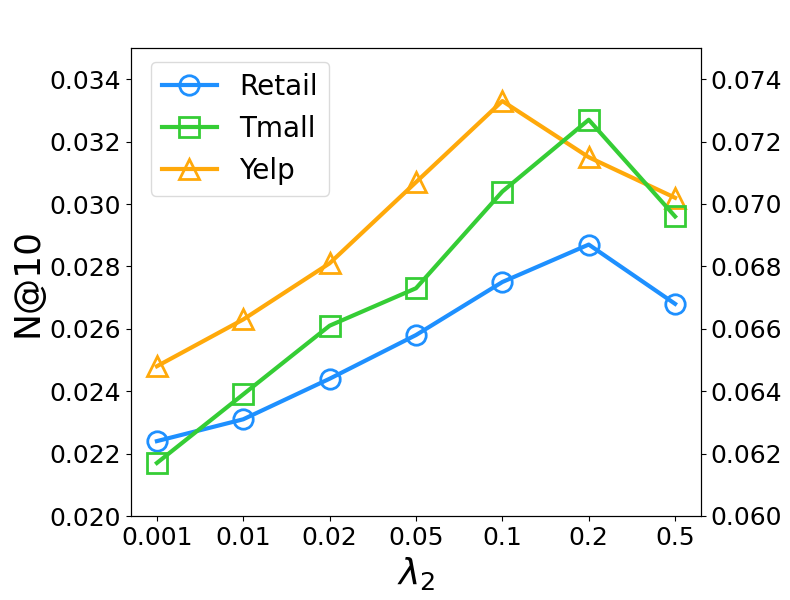}
        \label{fig:lambda_2}
    \end{subfigure}
    \vspace{-4mm}
    \caption{Effects of $\lambda_1$ and $\lambda_2$ in the final loss function.}
    \label{fig:lambda}
    \vspace{-5mm}
\end{figure}

\vspace{-4mm}
\subsection{Ablation Study}\label{abla}

To assess the effectiveness of each component in \model, we conduct further ablation studies on diverse variants. To be specific, we generate variants as follows:
\begin{itemize}[leftmargin=*]
    \item \textbf{\textit{w/o EBP}} removes explicit behavior pattern representation learner.
    \item \textbf{\textit{w/o MRE}} omits multiplex relation embedding aggregation.
    \item \textbf{\textit{w/o RCR}} removes relation chain representation learning.
    \item \textbf{\textit{w/o RCL}} excludes relation-based contrastive learning. 
    \item \textbf{\textit{w/o RCE}} excludes relation chain-aware encoder.
    \item \textbf{\textit{w/o FNL}} excludes the $\mathcal{L}_{BPR}^{final}$.
\end{itemize}

We present the recommendation performance results of variants on the three datasets in Fig.~\ref{fig:ablation}. Diverse variants show significantly poorer performance compared with the \model in terms of $R@10$ and $N@10$, which proves that our proposed five variants are effective and necessary. 
Among them, \textbf{\textit{w/o RCR}} performs the worst, which demonstrates the effectiveness of relation chains in extracting useful information from early relations to help learn embeddings of both users and items in later relations, as the latter relations in a relation chain are usually more revealing of a user's interactive preference for the item. \textbf{\textit{w/o RCL}} performs only better than \textbf{\textit{w/o RCR}}, which demonstrates the importance of contrastive learning in distinguishing between different relation types and enhancing the learning process of characterizing the target relation (\eg, `buy'). Meanwhile, based on the experimental results of \textbf{\textit{w/o MRE}}, it is important to take full advantage of the multiple user-item relations. 
\textbf{\textit{w/o RCE}}, on the other hand, learns the correlations and dependencies between different relations from a new view, especially considering the effect of auxiliary relations on the target relation in relation chains, and the experimental results demonstrate the significance of this component. 
The results of \textbf{\textit{w/o EBP}} once again confirm that explicit multi-relational behavior pattern representation learning is crucial and reasonable for multi-behavior recommendation tasks. 

\vspace{-4mm}
\subsection{Performance on Mitigating Data Sparsity}
In this section, we explain that the \model helps to mitigate the data sparsity problem. Experimental results with different interaction sparsity on three datasets are shown in Fig.~\ref{fig:sparsity_T}. In the experiments, we select several of the best-performing baselines, \ie, MB-CGCN, S-MBRec, CML, KMCLR, and BPHGNN. Specifically, we categorized users into six groups ("[0,4)", "[4,5)", "[5,6)", "[6,7)", "[7,10)", and "[10,60)") based on the interaction number (or called degrees) each user claims. 
The model performance measured by $R@10$ and $N@10$ metrics (shown on the right axis in Fig.~\ref{fig:sparsity_T}) is the average of all users in each group. 
The overall number of users in each group is depicted on the left axis in Fig.~\ref{fig:sparsity_T}. 
Based on the experimental results, we can conclude as follows: i) Increased user interactions improve recommendation accuracy for all methods. High-quality relational embeddings have more possibility to be obtained through adequate user-item interactions. ii) Considering that CML and KMCLR have better results in mitigating the data sparsity problem alone, our model consistently outperforms them, which confirms the fact that \model is indeed better at mitigating data sparsity. 

\begin{figure*}[t]
    \centering
    \captionsetup[subfigure]{font=footnotesize, labelfont=bf}
    \begin{subfigure}{0.3\textwidth}
        \includegraphics[width=\linewidth]{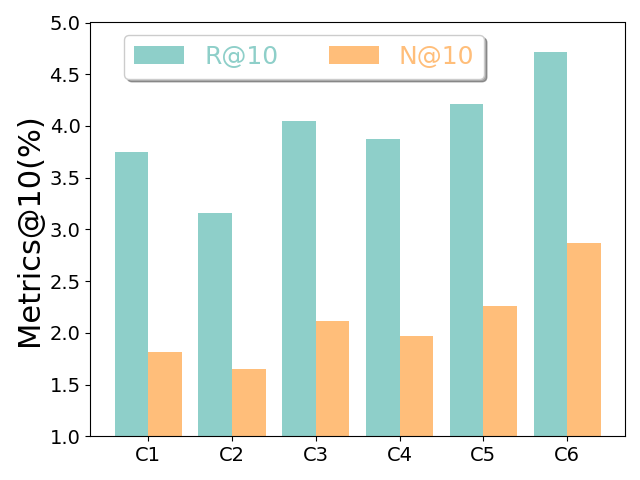}
        \caption{Retail}
        \label{fig:C_Retail}
    \end{subfigure}
    \hspace{-0.1cm}
    \begin{subfigure}{0.3\textwidth}
        \includegraphics[width=\linewidth]{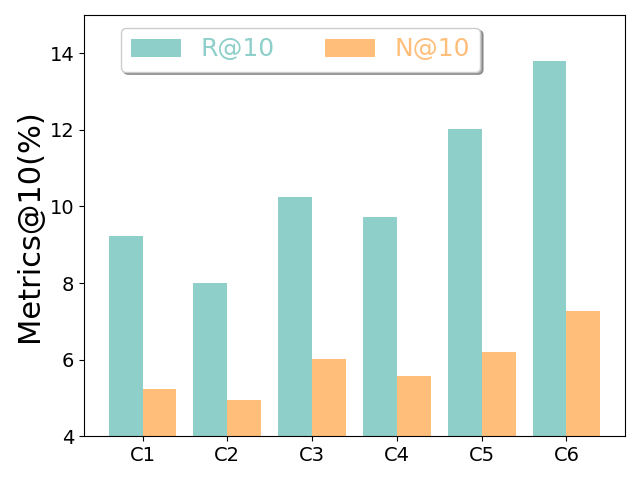}
        \caption{Tmall}
        \label{fig:C_Tmall}
    \end{subfigure}
    \hspace{-0.1cm}
    \begin{subfigure}{0.3\textwidth}
        \includegraphics[width=\linewidth]{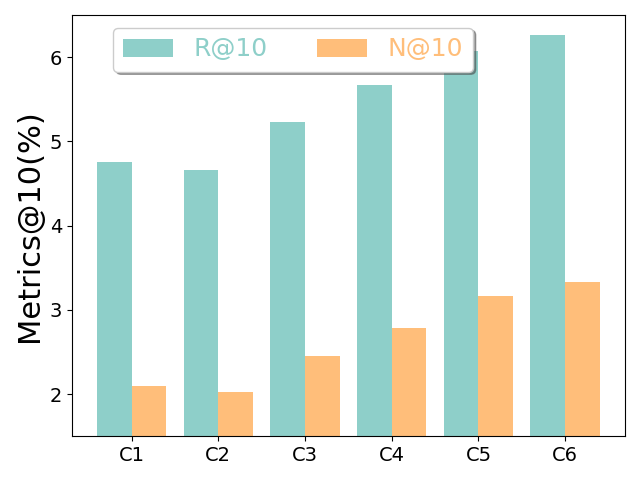}
        \caption{Yelp}
        \label{fig:C_Yelp}
    \end{subfigure}
    \caption{Effects of sequential orders in relation chains.}
    \label{fig:order}
    \vspace{-5mm}
\end{figure*}
\begin{figure*}[t]
    \centering
    \captionsetup[subfigure]{font=footnotesize, labelfont=bf}
    \begin{subfigure}{0.34\textwidth}
        \includegraphics[width=\linewidth]{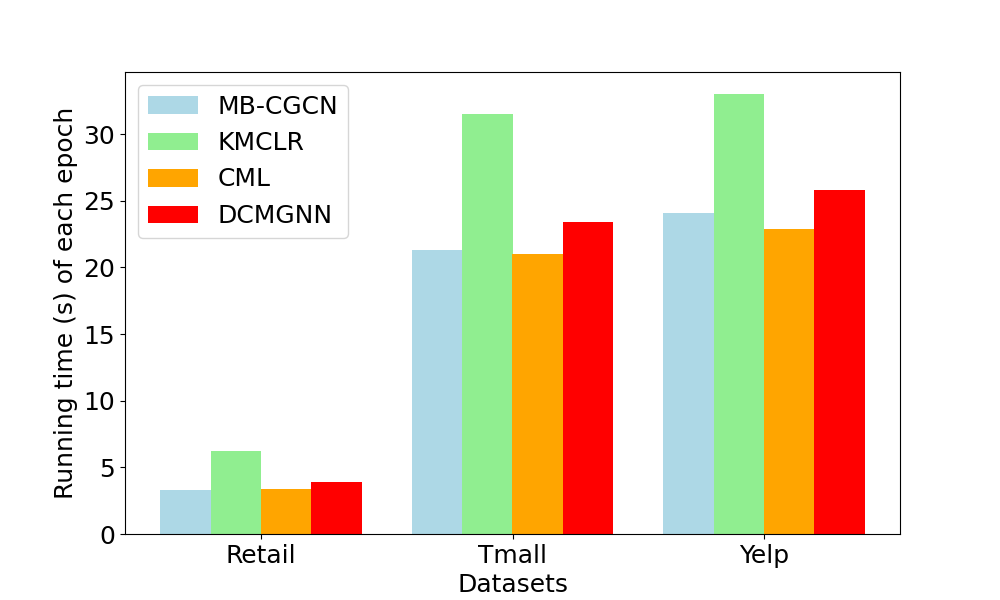}
        \caption{Running time of each epoch}
        \label{fig:et}
    \end{subfigure}
    \hspace{-0.6cm}
    \begin{subfigure}{0.34\textwidth}
        \includegraphics[width=\linewidth]{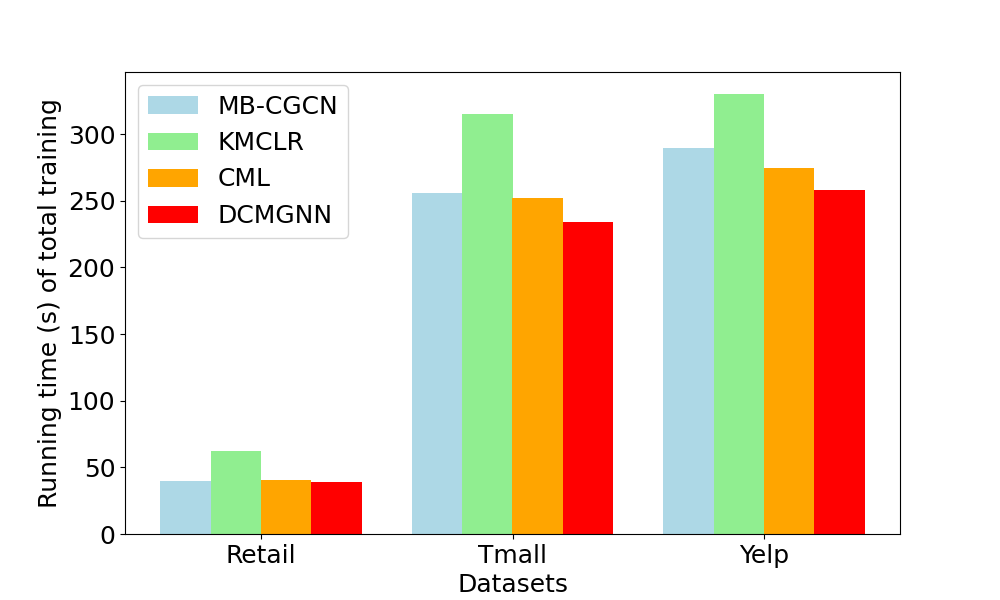}
        \caption{Running time of total training}
        \label{fig:tt}
    \end{subfigure}
    \hspace{-0.6cm}
    \begin{subfigure}{0.34\textwidth}
        \includegraphics[width=\linewidth]{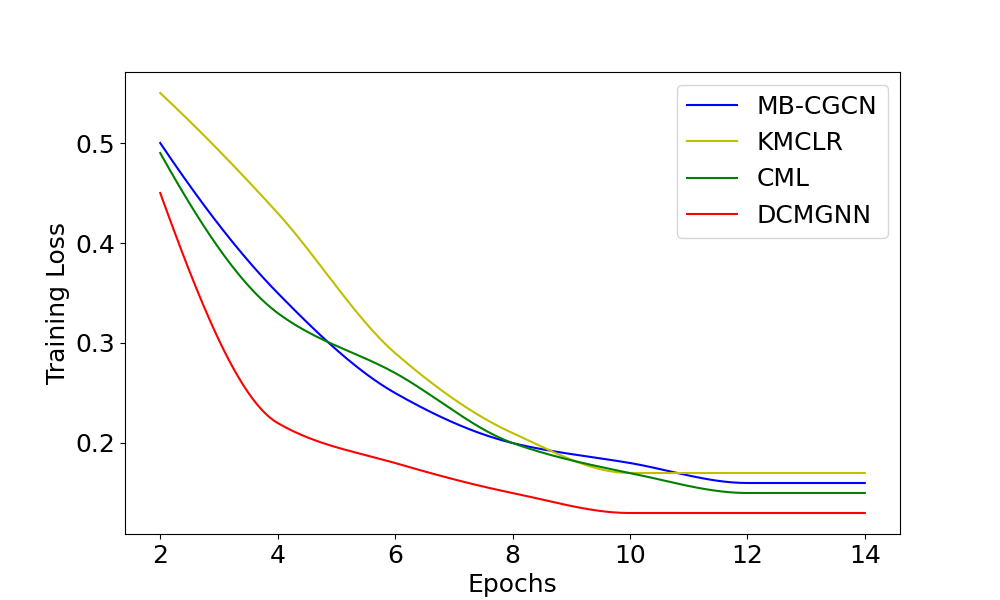}
        \caption{Training Loss}
        \label{fig:el}
    \end{subfigure}
    \caption{Training efficiency of methods.}
    \label{fig:te}
    \vspace{-5mm}
\end{figure*}
\vspace{-4mm}
\subsection{Parameter Sensitivity}
In order to assess the effect of embedding dimensionality $d$, the number of propagation layers $L$, and the order of the behavior chain on \model, we set up the following parameter sensitivity experiments, respectively:
\subsubsection{The Effect of Hidden Embedding Dimensionality}
As depicted in Fig.~\ref{fig:para1}, Retail and Yelp refer to the left axis, Tmall to the right axis. \model achieves the best performance when $d = 64$ on all three datasets. When $D$ is small, the performance of \model raises as the dimension increases to 64, after that, a higher dimension ($d>64$) may obviously lead to more computation cost and redundant information. 

\subsubsection{The Effect of Graph Propagation Layers} 
We can conclude from Fig.~\ref{fig:para1} that \model achieves better performance with the increase of the graph propagation layer ($L \leq 3$), it performs best when $L = 3$. This suggests that more message propagation layers within a certain range can capture more potential dependencies from higher-order neighbors. But continuing to superimpose more layers may bring noise effects or over-smoothing problems to the user and item representations. 

\subsubsection{The Effect of $\lambda_i$ in the final loss function}
To validate the effectiveness of various parts in the final loss function in Eq.~\ref{eq:final_loss}, we add the sensitivity analysis to hyperparameters $\lambda_1$ and $\lambda_2$, and the experimental results in $R@10$ and $N@10$ on three datasets are depicted in Fig.~\ref{fig:lambda}. The performance of \model exhibits a trend of initially increasing and then decreasing as $\lambda_1$ and $\lambda_2$ increase, indicating the effect of proposed relation chain-aware contrastive learning ($\mathcal{L}_{rcl}$) and the BPR loss of multi-relation node embedding ($\mathcal{L}_{BPR}$).

\subsubsection{The Effect of Relation Order}

For Tmall and Retail, the predefined relation chain order is $\langle View\xrightarrow{} Cart\xrightarrow{} Buy \rangle$, where `Buy' is the target relation. Specifically, different from Retail and Tmall, there exist four types of relations in Yelp, where `Dislike' and `Like' are mutually exclusive and do not appear simultaneously. Therefore, for Yelp, we define `Like' as the target relation, and the relation chain order is $\langle Neutral\xrightarrow{}Tips\xrightarrow{}Like \rangle$. We consider six containment order relation chains on Retail and Tmall datasets: $C1:\langle Buy\xrightarrow{} View\xrightarrow{} Cart \rangle$, $C2: \langle Buy\xrightarrow{} Cart\xrightarrow{} View \rangle$, $C3: \langle View\xrightarrow{} Buy\xrightarrow{} Cart \rangle$, $C4: \langle Cart\xrightarrow{} Buy\xrightarrow{} View \rangle$, $C5: \langle Cart\xrightarrow{} View\xrightarrow{} Buy \rangle$, and $C6: \langle View\xrightarrow{} Cart\xrightarrow{} Buy \rangle$. Meanwhile, we also consider six containment order relation chains on Yelp dataset: $C1:\langle Like\xrightarrow{} Neutral\xrightarrow{} Tips \rangle$, $C2: \langle Like\xrightarrow{} Tips\xrightarrow{} Neutral \rangle$, $C3: \langle Neutral\xrightarrow{} Like\xrightarrow{} Tips \rangle$, $C4: \langle Tips\xrightarrow{} Like\xrightarrow{} Neutral \rangle$, $C5: \langle Tips\xrightarrow{} Neutral\xrightarrow{} Like \rangle$, and $C6: \langle Neutral\xrightarrow{} Tips\xrightarrow{} Like \rangle$. The performance of the relation chains with different orders on three datasets is shown in Fig.~\ref{fig:order}. 
We find that the relation chains in which the target relation such as `Buy' or `Like' is the last relation shows the best performance, \ie, $C5$ and $C6$ are consistently better than other relation chains with different sequential orders. The order of relation chains affects user embeddings, as embeddings from previous relations influence those from subsequent ones. Thus, setting a reasonable relation order is essential, the chosen order in this work demonstrates significant advantages, highlighting the importance of relation order in recommendations.

\vspace{-2mm}
\subsection{Model Efficiency Analysis}

To comprehensively evaluate the time efficiency of \model, we conduct extensive experiments comparing both per-epoch training time and total convergence time across three benchmark datasets against the state-of-the-art models: MB-CGCN, KMCLR, and CML. The experimental results, as illustrated in Fig.~\ref{fig:te}, reveal several key observations: while MB-CGCN and CML demonstrate marginally faster processing speeds in single-epoch execution, \model exhibits superior convergence efficiency. This advantage is further corroborated by the training loss trajectory presented in Fig.~\ref{fig:el}, which demonstrates \model's accelerated convergence rate. Consequently, when considering the total training duration across all datasets, our DCMGNN achieves the most favorable time efficiency while simultaneously maintaining optimal performance metrics, outperforming all baseline models in both computational efficiency and recommendation accuracy.

\begin{figure}
    \begin{center}
    \includegraphics[width=0.5\textwidth]{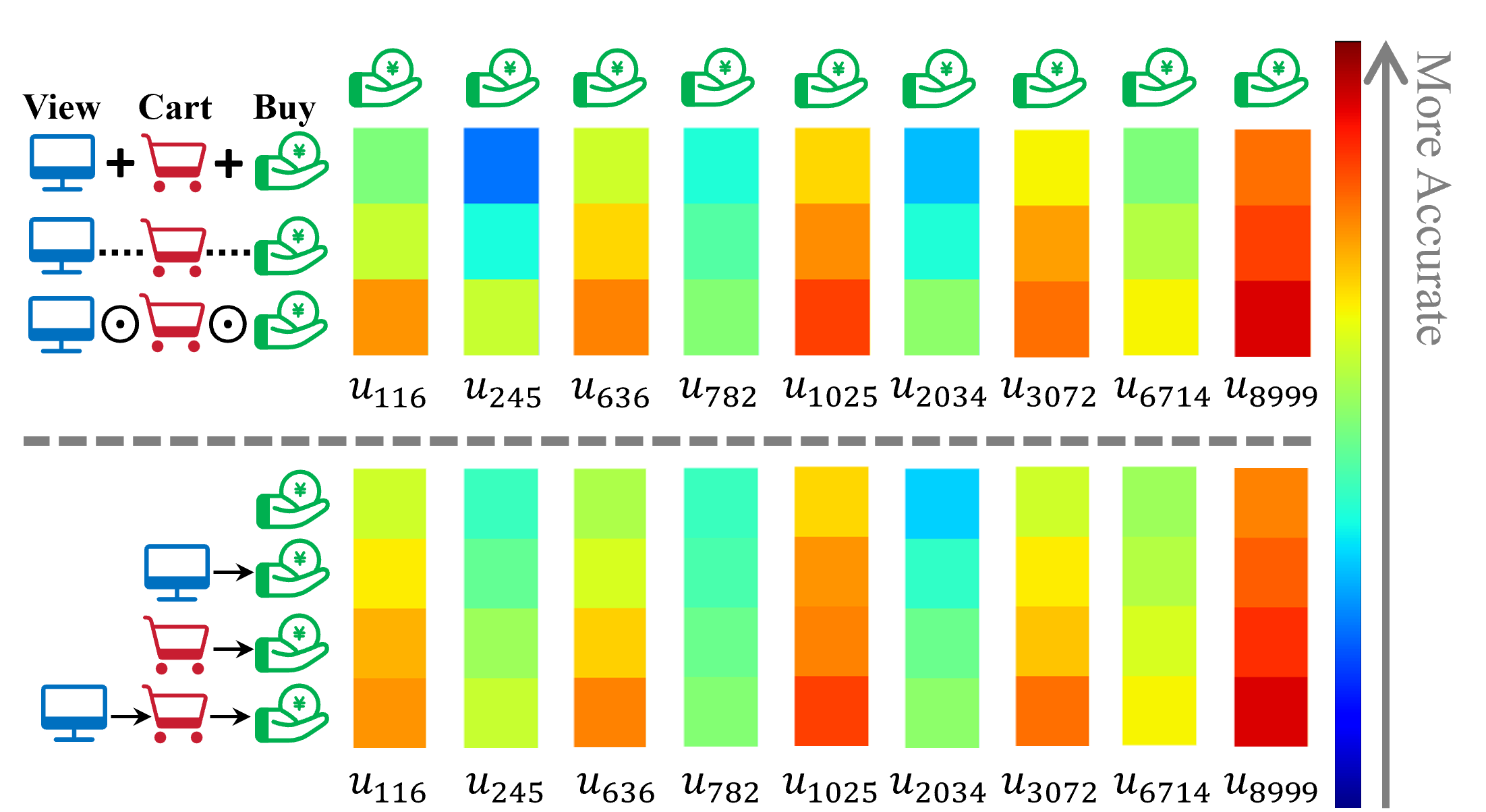}
    \caption{Case study on the interpretability of DCMGNN. Each small square in the heatmap represents the corresponding evaluated result of the sampled user in $NDCG@10$.}
    \label{fig:case_study} 
    \end{center}
    \vspace{-7mm}
\end{figure}

\vspace{-2mm}
\subsection{Case Study}\label{case_study}

We conduct case studies using real-world user examples in Tmall dataset to demonstrate the interpretability of \model. As shown in Fig.~\ref{fig:case_study}, the
upper and lower halves highlight the benefits of explicit behavior pattern modeling and implicit relation chain modeling on predicting the target relation.

To analyze the impact of different multi-behavior modeling methods, we compare three scenarios in the upper half of Fig.~\ref{fig:case_study}: (1) [`View'+`Cart'+`Buy'], a decoupled multi-behavior approach that models each individual relation and aggregates them, (2) [`View'$\cdot$$\cdot$$\cdot$$\cdot$`Cart'$\cdot$$\cdot$$\cdot$$\cdot$`Buy'], a meta-path-based approach that captures partial dependencies using predefined paths, and (3) [`View'$\odot$`Cart'$\odot$`Buy'], our proposed explicit behavior pattern modeling method. We observe that incorporating explicit behavior pattern modeling facilitates the acquisition of more accurate user representations by the model. The underlying rationale for this phenomenon is that explicit behavior pattern modeling enables the capture of user potential preferences within complex multi-behavior relationships.

We further examine the influences of different relation chains. The lower half of Fig.~\ref{fig:case_study} presents four configurations: $Buy$, $\langle View \to Buy \rangle$, $\langle Cart \to Buy \rangle$, and $\langle View \to Cart \to Buy \rangle$. Here, `View' reflects general interests, while `Cart' and `Buy' capture specific shopping intentions, with `Cart' serving as a transitional relation. Our used relation chain $\langle View \to Cart \to Buy \rangle$ achieves the best performance, demonstrating the critical role of relation chains in capturing sequential dependencies and the progressive evolution of user intent. Our framework effectively strengthens the diverse influences of auxiliary relations on the target relation, highlighting the importance of relation chains in capturing dependencies and correlations among different relations.

%% file: Tables/Datasets.tex
\begin{table}[t] 
\centering
\footnotesize
\caption{Statistical information of evaluation datasets.}
\begin{tabular}{m{0.85cm}<{\centering}|m{0.65cm}<{\centering}m{0.65cm}<{\centering}m{1.1cm}<{\centering}c} 
\toprule
    Datasets & Users & Items & Interactions & Relation Type \\
\midrule
    Retail & 2,174 & 30,113 & $9.7\times 10^4$ & \{View, Cart, Buy\} \\
    Tmall & 15,449 & 11,953 & $1.2\times 10^6$ & \{View, Cart, Buy\} \\
    Yelp & 19,800 & 22,734 & $1.4\times 10^6$ & \{Tips, Like, Neutral, Dislike\} \\
\bottomrule
\end{tabular}
\label{tab:datasets}
\vspace{-4mm}
\end{table}

%% file: Tables/Baselines.tex
\begin{table*}[t]
\centering
\caption{Performance comparison  across three real-world datasets in terms of $R@10$, $R@20$, $N@10$, and $N@20$. Marker * indicates the results are statistically significant (t-test with p-value $<$ 0.01).
}
\vspace{-2mm}
\small
\setlength{\tabcolsep}{1.5mm}{}	
\begin{tabular}{c|cccc|cccc|cccc} 
\toprule

\multirow{2}{*}{Method}  & \multicolumn{4}{c|}{Retail} & \multicolumn{4}{c|}{Tmall} & \multicolumn{4}{c}{Yelp} \\

 & $R@10$ & $R@20$ & $N@10$ & $N@20$ & $R@10$ & $R@20$ & $N@10$ & $N@20$ & $R@10$ & $R@20$ & $N@10$ & $N@20$ \\
\midrule

BPR & 0.0230 & 0.0316 & 0.0124 & 0.0144 & 0.0236 & 0.0311 & 0.0128 & 0.0152 & 0.0175 & 0.0287 & 0.0103 & 0.0129 \\
LightGCN & 0.0383 & 0.0438 & 0.0209 & 0.0233 & 0.0411 & 0.0546 & 0.0240 & 0.0266 & 0.0191 & 0.0302 & 0.0119 & 0.0144 \\
HCCF & 0.0396 & 0.0471 & 0.0214 & 0.0238 & 0.0424 & 0.0558 & 0.0242 & 0.0271 & 0.0189 & 0.0381 & 0.0133 & 0.0156 \\
DCCF & 0.0395 & 0.0476 & 0.0218 & 0.0239 & 0.0423 & 0.0561 & 0.0241 & 0.0279 & 0.0207 & 0.0405 & 0.0134 & 0.0161 \\
AutoCF & 0.0402 & 0.0477 & 0.0222 & 0.0238 & 0.0415 & 0.0525 & 0.0239 & 0.0277 & 0.0208 & 0.0399 & 0.0135 & 0.0174 \\
LightGCL & 0.0409 & 0.0489 & 0.0229 & 0.0242 & 0.0441 & 0.0597 & 0.0255 & 0.0289 & 0.0249 & 0.0451 & 0.0149 & 0.0189 \\
\midrule
RGCN & 0.0363 & 0.0446 & 0.0188 & 0.0204 & 0.0315 & 0.0426 & 0.0234 & 0.0275 & 0.0305 & 0.0537 & 0.0204 & 0.0237 \\
MBGCN & 0.0379 & 0.0457 & 0.0209 & 0.0227 & 0.0809 & 0.0991 & 0.0294 & 0.0350 & 0.0416 & 0.0634 & 0.0229 & 0.0258 \\
MB-HGCN & 0.0412 & 0.0472 & 0.0232 & 0.0258 & 0.1098 & 0.1783 & 0.0635 & 0.0852 & 0.0528 & 0.0698 & 0.0256 & 0.0274 \\
MB-CGCN & 0.0418 & 0.0492 & 0.0249 & 0.0253 & \underline{0.1233} & 0.2007 & 0.0660 & 0.0879 & 0.0573 & 0.0725 & 0.0285 & 0.0302 \\
CRGCN & 0.0411 & 0.0471 & 0.0232 & 0.0252 & 0.0855 & 0.1369 & 0.0539 & 0.0776 & 0.0561 & 0.0682 & 0.0237 & 0.0269 \\
BPHGNN & 0.0376 & 0.0449 & 0.0217 & 0.0245 & 0.0991 & 0.1806 & 0.0641 & 0.0836 & 0.0537 & 0.0705 & 0.0275 & 0.0289 \\
IMGCF & 0.0417 & 0.0485 & 0.0240 & 0.0253 & 0.1201 & 0.1888 & 0.0645 & 0.0842 & 0.0571 & 0.0729 & 0.0288 & 0.0313 \\
S-MBRec & 0.0386 & 0.0461 & 0.0234 & 0.0248 & 0.0877 & 0.1691 & 0.0642 & 0.0795 & 0.0559 & 0.0723 & 0.0287 & 0.0311 \\
\midrule
NMTR & 0.0372 & 0.0448 & 0.0198 & 0.0210 & 0.0682 & 0.0842 & 0.0273 & 0.0303 & 0.0397 & 0.0579 & 0.0215 & 0.0266 \\
MBCMN & 0.0405 & 0.0478 & 0.0217 & 0.0241 & 0.0857 & 0.1578 & 0.0607 & 0.0763 & 0.0537 & 0.0698 & 0.0273 & 0.0287 \\
MBA & 0.0428 & 0.0499 & 0.0251 & 0.0264 & 0.1230 & 0.2109 & 0.0665 & 0.0879 & 0.0578 & 0.0750 & 0.0296 & 0.0323 \\
HPMR & 0.0425 & 0.0500 & 0.0249 & 0.0261 & 0.1212 & 0.2085 & 0.0661 & 0.0870 & 0.0577 & 0.0741 & 0.0295 & 0.0321 \\
\midrule
HMG-CR & 0.0363 & 0.0446 & 0.0213 & 0.0234 & 0.0854 & 0.1546 & 0.0663 & 0.0714 & 0.0555 & 0.0707 & 0.0263 & 0.0275 \\
MMCLR & 0.0411 & 0.0489 & 0.0236 & 0.0260 & 0.1137 & 0.1769 & 0.0641 & 0.0828 & 0.0558 & 0.0720 & 0.0279 & 0.0308 \\
CML & 0.0428 & 0.0492 & 0.0247 & 0.0263 & 0.1203 & 0.2092 & \underline{0.0661} & 0.0852 & 0.0577 & 0.0745 & 0.0294 & \underline{0.0321} \\
KMCLR & \underline{0.0428} & \underline{0.0501} & \underline{0.0251} & \underline{0.0264} & 0.1215 & \underline{0.2107} & 0.0659 & \underline{0.0882} & \underline{0.0578} & \underline{0.0752} & \underline{0.0297} & 0.0319 \\
\midrule
\textbf{DCMGNN} & \textbf{0.0471*} & \textbf{0.0544*} & \textbf{0.0287*} & \textbf{0.0293*} & \textbf{0.1379*} & \textbf{0.2304*} & \textbf{0.0727*} & \textbf{0.0943*} & \textbf{0.0626*} & \textbf{0.0835*} & \textbf{0.0333*} & \textbf{0.0351*} \\
\midrule
\textit{Improvement} & 10.05\% & 8.58\% & 14.34\% & 10.98\% & 11.84\% & 9.35\% & 9.98\% & 6.92\% & 8.30\% & 10.91\% & 12.12\% & 9.35\% \\
\bottomrule
\end{tabular}
\label{tab:perform_compared}
\vspace{-4mm}
\end{table*}

%% file: Tables/Supply_Baselines.tex
\begin{table*}[h]
\centering
\caption{Performance comparison across three real-world datasets in terms of $R@5$, $R@40$, $N@5$, and $N@40$. Marker * indicates the results are statistically significant (t-test with p-value $<$ 0.01).}
\vspace{-2mm}
\small
\setlength{\tabcolsep}{1.5mm}{}	
\begin{tabular}{c|cccc|cccc|cccc}
\toprule

\multirow{2}{*}{Method}  & \multicolumn{4}{c|}{Retail} & \multicolumn{4}{c|}{Tmall} & \multicolumn{4}{c}{Yelp} \\

 & $R@5$ & $R@40$ & $N@5$ & $N@40$ & $R@5$ & $R@40$ & $N@5$ & $N@40$ & $R@5$ & $R@40$ & $N@5$ & $N@40$ \\
\midrule

BPR & 0.0212 & 0.0434 & 0.0102 & 0.0166 & 0.0216 & 0.0494 & 0.0112 & 0.0193 & 0.0134 & 0.0475 & 0.0095 & 0.0161 \\
LightGCN & 0.0357 & 0.0513 & 0.0201 & 0.0255 & 0.0341 & 0.0874 & 0.0214 & 0.0338 & 0.0154 & 0.0676 & 0.0101 & 0.0187 \\
HCCF & 0.0359 & 0.0537 & 0.0198 & 0.0258 & 0.0347 & 0.0899 & 0.0215 & 0.0346 & 0.0157 & 0.0687 & 0.0112 & 0.0196 \\
DCCF & 0.0358 & 0.0538 & 0.0212 & 0.0253 & 0.0344 & 0.0901 & 0.0215 & 0.0357 & 0.0151 & 0.0692 & 0.0120 & 0.0198 \\
AutoCF & 0.0361 & 0.0546 & 0.0212 & 0.0256 & 0.0370 & 0.0866 & 0.0209 & 0.0338 & 0.0155 & 0.0686 & 0.0119 & 0.0205 \\
LightGCL & 0.0366 & 0.0555 & 0.0223 & 0.0256 & 0.0401 & 0.0935 & 0.0227 & 0.0381 & 0.0178 & 0.0721 & 0.0137 & 0.0221 \\
\midrule
RGCN & 0.0305 & 0.0503 & 0.0134 & 0.0241 & 0.0134 & 0.0411 & 0.0111 & 0.0260 & 0.0169 & 0.0843 & 0.0185 & 0.0295 \\
MBGCN & 0.0359 & 0.0508 & 0.0195 & 0.0253 & 0.0389 & 0.1117 & 0.0231 & 0.0455 & 0.0183 & 0.0858 & 0.0201 & 0.0294 \\
MB-HGCN & 0.0372 & 0.0553 & 0.0207 & 0.0251 & 0.0415 & 0.1234 & 0.0565 & 0.0834 & 0.0187 & 0.0876 & 0.0209 & 0.0312 \\
MB-CGCN & 0.0381 & 0.0564 & 0.0223 & 0.0262 & \underline{0.0986} & \underline{0.3322} & 0.0564 & 0.1134 & 0.0225 & 0.1071 & 0.0241 & 0.0322 \\
CRGCN & 0.0375 & 0.0550 & 0.0218 & 0.0260 & 0.0744 & 0.2325 & 0.0403 & 0.0866 & 0.0197 & 0.0940 & 0.0217 & 0.0315 \\
BPHGNN & 0.0361 & 0.0537 & 0.0205 & 0.0261 & 0.0895 & 0.2703 & 0.0554 & 0.1089 & 0.0227 & 0.0976 & 0.0240 & 0.0332 \\
IMGCF & 0.0354 & 0.0567 & 0.0210 & 0.0259 & 0.0698 & 0.2583 & 0.0551 & 0.0857 & 0.0219 & 0.1087 & 0.0230 & 0.0315 \\
S-MBRec & 0.0372 & \underline{0.0593} & 0.0220 & 0.0269 & 0.0711 & 0.2593 & 0.0547 & 0.0880 & 0.0227 & 0.1135 & 0.0241 & 0.0337 \\
\midrule
NMTR & 0.0347 & 0.0515 & 0.0155 & 0.0249 & 0.0237 & 0.1034 & 0.0107 & 0.0383 & 0.0165 & 0.0824 & 0.0185 & 0.0305 \\
MBCMN & 0.0367 & 0.0529 & 0.0211 & 0.0255 & 0.0791 & 0.2529 & 0.0563 & 0.0855 & 0.0201 & 0.0879 & 0.0220 & 0.0318 \\
MBA & 0.0406 & 0.0580 & 0.0223 & 0.0270 & 0.0907 & 0.3063 & 0.0573 & 0.1227 & 0.0247 & 0.1150 & 0.0247 & 0.0345 \\
HPMR & 0.0406 & 0.0581 & 0.0222 & 0.0270 & 0.0909 & 0.3077 & 0.0572 & 0.1219 & 0.0245 & 0.1147 & 0.0249 & 0.0335 \\
\midrule
HMG-CR & 0.0341 & 0.0554 & 0.0197 & 0.0252 & 0.0606 & 0.2584 & 0.0551 & 0.0817 & 0.0213 & 0.0884 & 0.0223 & 0.0299 \\
MMCLR & 0.0352 & 0.0555 & 0.0209 & 0.0260 & 0.0690 & 0.2591 & 0.0549 & 0.0850 & 0.0218 & 0.1076 & 0.0231 & 0.0314 \\
CML & \underline{0.0408} & 0.0579 & 0.0224 & 0.0272 & 0.0915 & 0.3055 & 0.0573 & \underline{0.1272} & \underline{0.0249} & \underline{0.1157} & 0.0258 & 0.0342 \\
KMCLR & 0.0407 & 0.0577 & \underline{0.0228} & \underline{0.0273} & 0.0920 & 0.3235 & \underline{0.0574} & 0.1269 & 0.0243 & 0.1157 & \underline{0.0259} & \underline{0.0349} \\
\midrule
\textbf{DCMGNN} & \textbf{0.0446*} & \textbf{0.0624*} & \textbf{0.0248*} & \textbf{0.0298*} & \textbf{0.1099*} & \textbf{0.3466*} & \textbf{0.0618*} & \textbf{0.1368*} & \textbf{0.0275*} & \textbf{0.1237*} & \textbf{0.0282*} & \textbf{0.0382*} \\
\midrule
\textit{Improvement} & 9.31\% & 5.23\% & 8.77\% & 9.16\% & 11.46\% & 4.33\% & 7.67\% & 7.55\% & 10.44\% & 6.91\% & 8.88\% & 9.46\% \\
\bottomrule
\end{tabular}
\label{tab:supply_perform_compared}
\vspace{-2mm}
\end{table*}

%% file: Main/6_Conclusion.tex
\section{Conclusion}
In this paper, we propose a novel model called \model for multi-behavior recommendation task. The purpose of proposed components in \model is to learn feature embeddings of users and items from explicit and implicit perspectives. Especially, in the relation chain-aware encoder we reinforce the correlations and dependencies between different auxiliary relations and the target relation, which is one of the most important innovations of our model. Extensive experiments on three publicly available real-world datasets demonstrate that the proposed \model outperforms SOTA multi-behavior recommendation approaches with decent performance gains.